\documentclass[times,twocolumn,final,authoryear]{elsarticle}
\usepackage{jasr}
\usepackage{framed,multirow}

\usepackage{amssymb}
\usepackage{latexsym}
\usepackage{xfrac}

\usepackage[switch]{lineno}

\usepackage{url}
\usepackage{xcolor}
\definecolor{newcolor}{rgb}{.8,.349,.1}

\usepackage{diagbox}

\usepackage[citebordercolor=white]{hyperref}

\journal{Advances in Space Research}

\begin{document}

\verso{Eduardo Maria Polli  \textit{et al}}

\begin{frontmatter}

\title{Analytical model for collision probability assessments with large satellite constellations}

\author[1]{Eduardo Maria \snm{Polli}}
\ead{eduardo.polli94@gmail.com}
\author[1]{Juan Luis \snm{Gonzalo}}
\ead{juanluis.gonzalo@polimi.it}
\author[1]{Camilla \snm{Colombo}}
\ead{camilla.colombo@polimi.it}

\address[1]{Politecnico di Milano, Via La Masa 34, 20158 Milan, Italy}

\begin{abstract}
At a time when space debris are already a growing issue in the space sector, the deployment of large constellations, made of hundreds to thousands of satellites, is of concern from an environmental point of view. In the next decade, the space sector will undergo a considerable change as the population of active satellites is about to quintuple. This scenario will pose new challenges regarding space traffic management, generating the demand for more powerful and efficient analysis tools. \\
In this study, an analytical model for collision probability assessments between de-orbiting or injecting space objects and satellite constellations is presented. Considering the first to be subjected to a continuous tangential acceleration, its spiraling motion would result in a series of close approaches in the proximity of a constellation. The mathematical description of the crossing dynamics relies on the assumption of circular orbits and independent collision probabilities, but does not require to propagate the satellites' orbit and it is suitable also for elliptical crossing orbits. The statistical model presented in the current work constitutes an efficient tool for the evaluation of the mean collision probability related to this type of events. A comparison with a conventional propagation method has been performed for validation purposes. \\
The statistical model has been used to assess the risk connected to constellation's satellites replacement, once they have reached their programmed End-of-Life. The environmental impact of the full replacement of $12$ approved constellations is analysed by means of average collision probability. In particular, it is shown that the key features for space exploitation sustainability are the maximum propulsion available from the thruster, the selection of an optimal crossing orbit and the true anomaly phases between constellations' and crossing satellites. The consequences of an in-orbit collision are also investigated by assessing the collision risk generated by the formation of a debris cloud. \\
The results corroborate the need for international standards for space traffic management as an exponentially increasing satellites population could trigger a chain reaction of collisions, making LEO inaccessible for decades.
\end{abstract}


\end{frontmatter}


\section{Introduction}

The exponential growth the private space sector experienced in the last decade led to an improvement of the engineering knowledge and, eventually, to a reduction of the cost of payload launch per kg. This, along with the more widespread use of cellphones and Internet among the population, allowed the realization of telecommunications satellite constellations. Space-based world-wide telecommunications could play a key role as the cost of the infrastructures related to on-ground Internet connections has left many rural and underdeveloped regions unserved. Indeed, large constellations appear to be the most suitable option to bridge the digital divide, which is one of the UN Sustainable Development Goals. Moreover, for the first time, near global coverage would be available with only one provider. \\
In particular, SpaceX \citep{bib_starlink}, OneWeb \citep{bib_oneweb}, Telesat \citep{bib_telesat} and Amazon \citep{bib_kuiper} plan to deploy large constellations composed by thousands of satellites within the next years, all four projects being already approved by the Federal Communications Commission (FCC), the authority responsible for spectrum usage and for launch approval, along with the Federal Aviation Administration (FAA), in the United States. \\
As most Internet services requrie fast links, satellite constellations for telecommunication purposes are going to be deployed in Low Earth Orbit (LEO). The main advantage of LEO constellations with respect to higher-altitude satellite telecommunication systems, is the low latency due to the shorter satellite-Earth distances involved. The latency of LEO communications can be more than ten times lower than the one from Geostationary orbits. Furthermore, the smaller footprint of each antenna allows to reduce the size of the hardware components and, consequently, of the thruster, thus reducing the manufacturing costs of each satellite. Nevertheless, cheaper satellites are typically characterised by a shorter life-time cycle and, because the worst downside of LEO constellations rests in the number of satellites needed for (near) global coverage, this feature would also increase satellite traffic in LEO. \\
Constellation satellites have to be replaced once they reach their End-of-Life (EoL). Considering an average operational phase of 5 years as done in \citet{bib_radtke}, more than $10$ replacements per day would be necessary for the $18,348$ already approved constellation's satellites considered in the current work. Furthermore, this will generate a two-way traffic, since a replacement involves the disposal of the EoL satellite and the injection of the new one. \\
According to the Union of Concerned Scientists satellite database \citep{bib_uocs} updated on January $1^{\text{st}}$ $2022$, there are currently $4,852$ active satellites orbiting around Earth, $4,078$ of which are in LEO. To these must be added around $3,000$ satellites that are either dead or currently on their de-orbiting phase and more than $25,000$ objects classified as space debris. In this environment, the injection of more than $18,000$ satellites in LEO has already been approved, mostly by the FCC \citep{bib_tesi}. \\
The increasing number of objects orbiting around Earth has motivated different analyses of the environmental impact of satellite constellations, mostly based on the interaction with space debris. In $1998$, \cite{bib_rossi} developed a semi-deterministic model for the estimation of the number of collisions between satellite constellations and space debris based on a series of Monte Carlo (MC) simulations, also considering different traffic scenarios. In \citet{bib_rossi17}, the environmental impact of satellite constellations was analysed by mean of the Criticality of Constellation Index, or CCI. Different physical properties of the satellites, operational capabilities and constellation designs were considered for the $68$ MC simulations performed. It was found that, among the investigated ones, the most critical parameters are: mass, area, and failure probability during the operational phase. \\
Other investigations rely on the ESA-MASTER software for the evaluation of the debris flux during the different phases of a constellation life cycle. The debris flux on the large constellations which most probably will be completed first, namely Starlink and OneWeb, was analysed accurately in \citet{bib_radtke} and \citet{bib_lemay}. As an example, considering Business-As-Usual scenario, the probability of a catastrophic collision over the five years operational phase between a non-trackable debris and the high-altitude shells of the Starlink constellation evaluated in \citet{bib_lemay} is $10.20 \ \%$. \\
The general approach for this type of risk assessments involves the characterization of the space debris environment in which the constellation is deployed, typically through a subdivision into volume elements. The collision probability is obtained from the number of collisions, which are statistically estimated. More preicse algorithms ca be used assess the probability of a collision between two approaching objects, whose nominal orbits and position errors are known. Most known algorithms are \citet{bib_chan}, \citet{bib_alfano}, \citet{bib_foster} and \citet{bib_patera}. More recent ones, such as \citet{bib_serra} and \citet{bib_garcia}, include analytical solutions in case of anisotropic distributions. All these methods are based on a three-dimensional Gaussian distribution of the satellites’ positions and the collision probability is evaluated as the integral of the probability density function (PDF). \\
The objective of this study is to develop an analytical model that combines the efficiency of a statistical approach with the precision of a collision probability algorithm, in order to evaluate the mean collision probability related to a shell-crossing event. This type of event involves the passage of a de-orbiting or injecting satellite, under continuous tangential thrust, which has to cross the thin region of space populated by constellation’s satellites, in order to reach its target destination. Given the large satellite population, most collision probability algorithms require large computational efforts to assess the risk related to an orbit-raising or -decreasing manoeuvre. On the other hand, the statistical model hereby presented is an efficient tool for preliminary mission design involving continuous tangential thrust. Moreover, the same model can be used to assess the collision probability with space debris, offering an alternative to the number of estimated collisions, which is currently used in software such as ESA-MASTER. Finally, given the statistical nature of the approach, this could represent a useful tool to assess the criticality of satellite injection and disposal, in a specific space environment. The model would be particularly convenient in this context, as it offers a risk assessment which does not depend on the position uncertainties of the satellites. \\
The injection of satellites can be divided into low- and high-thrust orbit raising. For LEO satellites, electric propulsion is typically adopted. This results in a continuous tangential low-thrust \citep{bib_pollard1} such that the satellite reaches its final orbit through a spiraling trajectory, having its semi-major axis continuously increasing. On the other hand, LEO satellites disposal is mainly performed via perigee decrease \citep{bib_pollard2}. This consists in lowering the perigee and letting the greater atmospheric drag of the lower altitudes slow down the satellite until re-entry. In this study, continuous tangential low-thrusting has been considered for both injection and disposal, as it also includes passive re-entry. \\
The mathematical description of the satellite constellation refers to the Walker constellation \citep{bib_walker}. This is a highly symmetrical design for satellite constellations since it involves a set of equally inclined, equally populated and equally spaced in RAAN circular orbits, all having the same semi-major axis. This is the most widely adopted design for large constellations and, due to its high degree of symmetry, allows for great simplifications for the evaluation of the average collision probability. \\
The environmental impact of satellite constellations is investigated by means of two different analyses: satellite constellation replacement and consequences of an in-orbit catastrophic collision. The former involves the evaluation of the mean collision probability between EoL de-orbiting satellites and new injecting satellites with both space debris and lower-altitude constellations. The latter assesses the collision probability between the cloud of fragments generated by an in-orbit collision, and lower-altitude constellations. \\
The reminder of the manuscript is organised as follows. In Section \ref{model_sec} a new model for estimation of collision probability averaged over every possible true anomaly phase between crossing and constellations satellites is presented. A simplified propagation model is developed in Section \ref{moidprop_sec}. This can be used to assess, in a highly efficient way, the collision probability profile over the initial true anomaly phase between crossing and constellation's satellites, when both orbits are circular. Section \ref{valid_sec} presents the validation of both the statistical and the simplified propagation model. Constellations replacements and consequences of a in-orbit catastrophic collision are analysed, respectively, in Sections \ref{result1_sec} and \ref{result2_sec}, using the statistical model developed in Section \ref{model_sec}. Section \ref{conclusion_sec} presents the conclusion.
  
\section{Model}
\label{model_sec}

This section proposes a simplified statistical model specifically designed for the evaluation of the collision probability between a crossing satellite and all the satellites of a constellation shell, assuming every orbit to be circular. \\
The model is based on the assumption that all the $\overline{N}_{CA}$ close approaches between the crossing satellite and one constellation's satellite, share the same probability $\overline{P}$ that a collision will occur. Under this assumption, and considering the probabilities of different close approaches as independent, the overall collision probability $P_{satellite}$ between a constellation's and a crossing satellite is:
\begin{equation}
    P_{satellite} = 1 - \left( 1 - \overline{P} \right)^{\overline{N}_{CA}}
    \label{p_plane_start}
\end{equation}
In the following, the technique for the evaluation of $\overline{P}$ and $\overline{N}_{CA}$ will be explained.

\subsection{Collision probability}

In this section are presented the key ideas and the implementation of Chan's algorithm, for the evaluation of the collision probability between two orbiting objects. The formula obtained is then simplified by considering the orbits of both objects to be circular. \\
Let $S_1$ and $S_2$ be two satellites on their respective nominal orbits $\kappa_1$ and $\kappa_2$, with covariance matrices at the time of closest approach (TCA) $\boldsymbol{\zeta}_1$ and $\boldsymbol{\zeta}_2$. Then, the collision probability between these two satellites can be evaluated using Chan's algorithm \citep{bib_chan}. \\
Let $[\mathbf{r}_1^{N},\mathbf{v}_1^{N}]$ and $[\mathbf{r}_2^{N},\mathbf{v}_2^{N}]$ be the Cartesian coordinates in the Earth-Centred Inertial (ECI) frame at the TCA of $S_1$ and $S_2$, respectively. Then, the axes of the encounter reference frame $\Pi$ centred at $S_2$ are defined at the TCA as follows:
\begin{itemize}
    \item $x-$axis along the relative position vector $\Delta \mathbf{r}^N = \mathbf{r}_1^{N} - \mathbf{r}_2^{N}$
    \item $y-$axis along the relative velocity vector $\Delta \mathbf{v}^N = \mathbf{v}_1^N - \mathbf{v}_2^N$
    \item $z-$axis completing the set of principal axes
\end{itemize}
Note that $x$ and $y$ axes are perpendicular at TCA, as the miss distance $|\Delta \mathbf{r}^N|$ is minimum if and only if $\Delta \mathbf{r}^N \perp \Delta \mathbf{v}^N$. Moreover, recalling that the encounter plane is perpendicular to the relative velocity vector, this is the $x$-$z$ plane. \\
The position of $S_1$ in $\Pi$ is expressed as $\boldsymbol{r_p}=(d,0,0)$, where $d$ is the nominal miss distance at the TCA. Modelling the satellites as spheres of radii $R_1$ and $R_2$, the collision occurs when the relative distance between $S_1$ and $S_2$ is smaller than the radius of the combined hard sphere of the bodies, i.e. $r_a = R_1 + R_2$. \\
Under the hypothesis of uncorrelated measurements, the combined covariance matrix of the satellites can be computed as $\boldsymbol{\zeta} = \boldsymbol{\zeta}_1 + \boldsymbol{\zeta}_2$ and it can be expressed as:
\begin{equation}
    \boldsymbol{\zeta} = 
    \begin{bmatrix}
    \sigma_x^2 & \rho_{xy} \sigma_x \sigma_y & \rho_{xz} \sigma_x \sigma_z \\
    \rho_{xy} \sigma_x \sigma_y & \sigma_y^2 & \rho_{yz} \sigma_y \sigma_z \\
    \rho_{xz} \sigma_x \sigma_z & \rho_{yz} \sigma_y \sigma_z & \sigma_z^2
    \end{bmatrix}
\end{equation}
The PDF of a collision is modeled inside the $x$-$z$ encounter plane. Moreover, the diagonal form of $\boldsymbol{\zeta}'$ of the combined covariance matrix $\boldsymbol{\zeta}$ can be obtained by moving through principal axes through a rotation around the $y-$axis. The required rotation angle $\gamma$ is given by:
\begin{equation}
    \gamma = \frac{1}{2} \tan^{-1} \left( \frac{2 \rho_{xz} \sigma_x \sigma_z}{\sigma_x^2 - \sigma_z^2} \right)
\end{equation}
and, on the encounter plane, it would be equal to:
\begin{equation}
    \boldsymbol{\zeta}' = \boldsymbol{R}_\gamma \cdot \boldsymbol{\zeta} \cdot \boldsymbol{R}_\gamma^T = \begin{bmatrix} \sigma_{x'}^2&0\\
    0&\sigma_{z'}^2 \end{bmatrix}
    \label{chan2}
\end{equation}
where $\boldsymbol{R}_\gamma$ is the rotation matrix of angle $\gamma$, around the $y-$ axis.\\
The collision probability $P$ between $S_1$ and $S_2$ can be evaluated as \citep{bib_chan}:
\begin{equation}
    P = \text{e}^{-\tfrac{V}{2}} {\displaystyle \sum_{j=0}^{\infty} \frac{V^j}{2^j j!} \left[ 1 - \text{e}^{-\tfrac{U}{2}} {\displaystyle \sum_{k=0}^{j} \frac{U^k}{2^k k!}} \right]}
    \label{chan}
\end{equation}
where:
\begin{equation}
    U = \left(\frac{r_a^2}{\sigma_{x'} \sigma_{z'}} \right)
    \ \ \ \ \ \ \ \ \ \ \ \ \ \ \ \ \ \ 
    V = \left( \frac{r_p''}{\sigma_{z'}} \right)^2
    \ \ \ \ \ 
\end{equation}
Let $r_p'$ be the miss distance vector in the encounter frame, then $r_p''$ is obtained by scaling $r_p'$ along the $x$ direction, such that the constant PDF ellipses turn into circles and the distribution becomes isotropic. It follows that:
\begin{equation}
    r_p''^2 = \left(\tfrac{\sigma_{z'}}{\sigma_{x'}} d \cos \gamma \right)^2 +  \left( d \sin \gamma \right)^2
\end{equation}
In \citet{bib_chan} it is stated that, based on empirical evidences, the axes of a satellite error ellipsoid are almost aligned with the axes of its RSW reference frame. It follows that the covariance matrix of a satellite is almost diagonal in its RSW frame, this being defined as:
\begin{itemize}
    \item $R-$axis along the radial direction, pointing towards the centre of Earth
    \item $S-$axis along the along-track direction, parallel to the velocity vector for circular orbits
    \item $W-$axis along the cross-track direction, completing the set of axes
\end{itemize}
If the covariance matrices $\boldsymbol{\zeta}_1^{RSW}$ and $\boldsymbol{\zeta}_2^{RSW}$ of, respectively, $S_1$ and $S_2$ are considered to be diagonal in their respective RSW reference frames \citep{bib_chan}, a great simplification of the problem is presented in \citet{bib_bai}. The following encounter frame $\Pi^{\circ}$ is first defined:
\begin{itemize}
    \item $\hat{x}-$axis along the radial direction of $S_2$
    \item $\hat{y}-$axis along the relative velocity between $S_1$ and $S_2$
    \item $\hat{z}-$axis completing the set of principal axes
\end{itemize}
The position of $S_1$ in $\Pi^{\circ}$ becomes:
\begin{equation}
    \mathbf{r}_1^{\circ} = [\mu_{\hat{x}},0,\mu_{\hat{z}}]
    \label{mu_x_mu_z_def}
\end{equation}
where $d = \sqrt{\mu_{\hat{x}}^2 + \mu_{\hat{z}}^2}$ is the nominal miss distance at the TCA. \\
Let $\boldsymbol{\zeta}_1^{RSW}$ and $\boldsymbol{\zeta}_2^{RSW}$ be defined as:
\begin{equation}
    \boldsymbol{\zeta}_1^{RSW} = \begin{bmatrix}
    \sigma_{1_R}^2 & 0 & 0 \\
    0 & \sigma_{1_S}^2 & 0 \\
    0 & 0 & \sigma_{1_W}^2 \\
    \end{bmatrix}
    \ \ \
    \boldsymbol{\zeta}_2^{RSW} = \begin{bmatrix}
    \sigma_{2_R}^2 & 0 & 0 \\
    0 & \sigma_{2_S}^2 & 0 \\
    0 & 0 & \sigma_{2_W}^2 \\
    \end{bmatrix}
    \label{zeta_rsw_def}
\end{equation}
then, the resulting combined covariance matrix $\boldsymbol{\zeta}^{\circ}$ on the encounter plane becomes:
\begin{equation}
    \boldsymbol{\zeta}^{\circ} = \begin{bmatrix} \sigma_{\hat{x}}^2&0\\0&\sigma_{\hat{z}}^2 \end{bmatrix}
    \label{chan3}
\end{equation}
where:
\begin{subequations}
\begin{align}
    \sigma_{\hat{x}}^2 & = \sigma_{1_R}^2 + \sigma_{2_R}^2
    \label{sigma_x_bai}
    \\
    \sigma_{\hat{z}}^2 & = (\sigma_{1_S}^2 + \sigma_{2_S}^2) \cos^2\left( \tfrac{\varphi}{2} \right) + (\sigma_{1_W}^2 + \sigma_{2_W}^2) \sin^2\left( \tfrac{\varphi}{2} \right)
    \label{sigma_z_bai}
\end{align}
\label{sigma_bai}%
\end{subequations}
The angle $\varphi$ is the angle between the angular momenta of the orbits $\kappa_1$ and $\kappa_2$. \\
The collision probability can finally be evaluated as:
\begin{equation}
    P = e^{-\tfrac{\hat{V}}{2}} {\displaystyle \sum_{j=0}^{\infty} \frac{\hat{V}^j}{2^j j!} \left[ 1 - e^{-\tfrac{\hat{U}}{2}} {\displaystyle \sum_{k=0}^{j} \frac{\hat{U}^k}{2^k k!}} \right]}
    \label{bai}
\end{equation}
where:
\begin{equation}
    \hat{U} = \left( \frac{r_a^2}{\sigma_{\hat{x}} \sigma_{\hat{z}}} \right)
    \ \ \ \ \ \ \ \ \ \ \ \ \ \ \ \ \ \ \ \ \ \ 
    \hat{V} = \left( \frac{\mu_{\hat{x}}^2}{\sigma_{\hat{x}}^2} + \frac{\mu_{\hat{z}}^2}{\sigma_{\hat{z}}^2} \right)
\end{equation}
Until now, three approximations have been introduced into the model:
\begin{itemize}
    \item satellites are modeled as spheres
    \item PDF is modified to be isotropic
    \item error ellipsoids are aligned with RSW axes
\end{itemize}
For typical space applications, the errors arising from these simplifications are, in general, negligible. 

\subsection{Satellite dynamics}

During the injection and active disposal phases the thrust is assumed to be tangential to the trajectory. This method is called tangential thrusting and represents the in-plane optimal control law for semi-major axis variations \citep{bib_simeng}. Considering both atmospheric drag and thruster propulsion, the dynamics of the re-entering or newly deployed constellation's satellite are fully described through the expression of the semi-major axis variation:
\begin{equation}
    \dot a = \dot a_{drag} + \dot a_{thrust}
    \label{a_dot_intro}
\end{equation}
where $\dot a_{drag}$ and $\dot a_{thrust}$ represent the semi-major axis rate of change due to air drag and tangential low-thrust, respectively. \\
Using the Gauss form of the variational equation of semi-major axis reported in \citet{bib_battin}, the dynamics of a satellite on a circular orbit can be defined as:
\begin{equation}
    \dot a = 2 \sqrt{\frac{a^3}{\mu}} a_{\theta}
    \label{Gauss}
\end{equation}
where:
\begin{itemize}
    \item $a$ is the semi-major axis
    \item $\mu$ is Earth standard gravitational parameter
    \item $a_{\theta}$ is the transversal component of the perturbative acceleration vector $\mathbf{a}_p$
\end{itemize}
If Earth rotation is neglected, drag is always tangential to the satellite velocity and both forces act along the same direction. Nonetheless, the rotation of the atmosphere can still be considered for the estimation of $\dot a_{drag}$ as follows:
\begin{equation}
    \dot a_{drag} = -\sqrt{\mu a} \frac{\rho C_D A}{M} \left(1 - \frac{w_\oplus \cos i}{n} \right)^2
    \label{adot_drag}
\end{equation}
where:
\begin{itemize}
    \item $\rho$ is the air density
    \item $C_D$ is the drag coefficient
     \item $A$ is the cross-sectional area of the spacecraft
    \item $M$ is the mass of the spacecraft
    \item $w_\oplus = \tfrac{2 \pi}{\textrm{day}}$ is the atmosphere angular velocity
    \item $i$ is the inclination of the the orbit
    \item $n = \sqrt{\frac{\mu}{a^3}}$ is the mean motion of the spacecraft
\end{itemize}
It must be emphasised that drag acceleration, in general, is not tangential to the satellite trajectory, due to the rotation of the atmosphere. Nevertheless, given the slow rate at which Earth rotates, these effects are considered to be negligible. \\
The dynamics of the thruster are described as:
\begin{equation}
    \dot a_{thrust} = \pm 4 \sqrt{\frac{a^3}{\mu}} \frac{\eta_t P_t}{M g_0 I_{sp}}
    \label{adot_thrust}
\end{equation}
where:
\begin{itemize}
    \item $\eta_t$ is the efficiency of the thruster
    \item $P_t$ is the input power of the thruster
    \item $g_0$ is the standard gravity
    \item $I_{sp}$ is the specific impulse of the thruster
\end{itemize}
The derivation of Eq.s (\ref{adot_drag}) and (\ref{adot_thrust}) are shown in \citet{bib_tesi} and \citet{bib_simeng}.

\subsection{Collision probability distribution}

\begin{figure}[t]
    \centering
    \includegraphics[width=\linewidth]{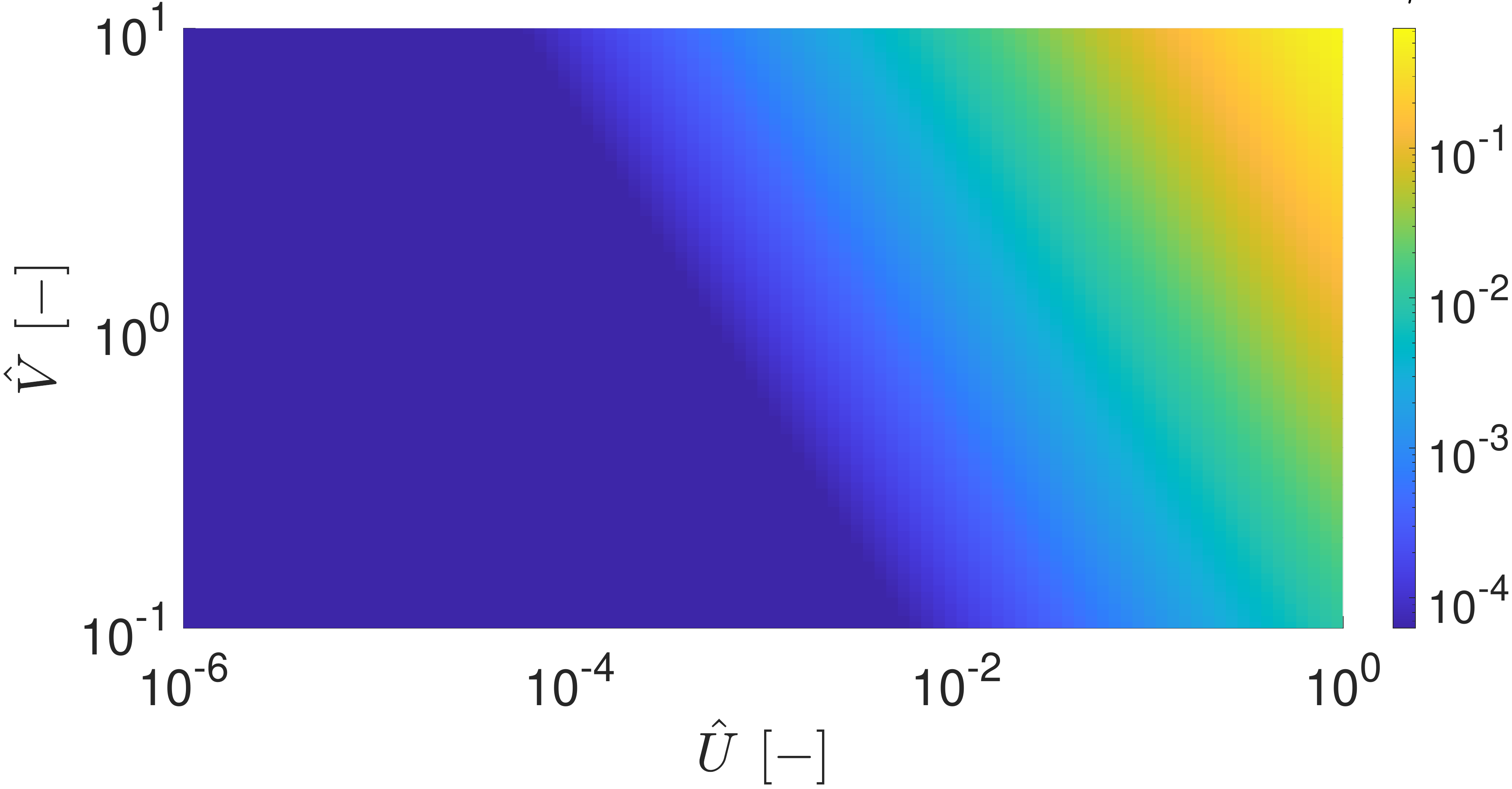}
    \caption{Relative error $\varepsilon_r$ between the $4^{\textrm{th}}$ term and $1^{\textrm{st}}$ term truncation of Eq. (\ref{bai}) for different values of $\hat{U}$ and $\hat{V}$.}
    \label{S1_max}
\end{figure}

The major benefit of Eq.s (\ref{chan}) and (\ref{bai}) is that the series converges rapidly \citep{bib_chan}. For typical satellites applications, it can be assumed that $\hat{U} < 10^{-4}$ and, as shown in Fig. \ref{S1_max}, the truncation of Eq. (\ref{bai}) at the first term still yields accurate results.
\\
The truncation at the first term of Eq. (\ref{bai}) is expressed through the following bivariate distribution:
\begin{equation}
    P \left( \mu_{\hat{x}}, \mu_{\hat{z}} \right) = P_{\odot} \exp \left( -\frac{1}{2} \frac{\mu_{\hat{x}}^2}{\sigma_{\hat{x}}^2} \right) \exp \left( -\frac{1}{2} \frac{\mu_{\hat{z}}^2}{\sigma_{\hat{z}}^2} \right)
    \label{prob_mu}
\end{equation}
where:
\begin{equation}
    P_\odot = 1 - \exp \left( - \frac{r_a^2}{ 2 \sigma_{\hat{x}} \sigma_{\hat{z}} } \right)
    \label{bai_moid1}
\end{equation}
does not depend on the miss distance, but rather on the geometrical and orbital properties of the satellites. $P_\odot$ is the value of the maximum collision probability between two approaching satellites, which does occur when the nominal miss distance is null. \\

\subsection{Nominal closest approach}
\label{NCA_sec}

\begin{figure}[t]
    \centering
    \includegraphics[width=\linewidth]{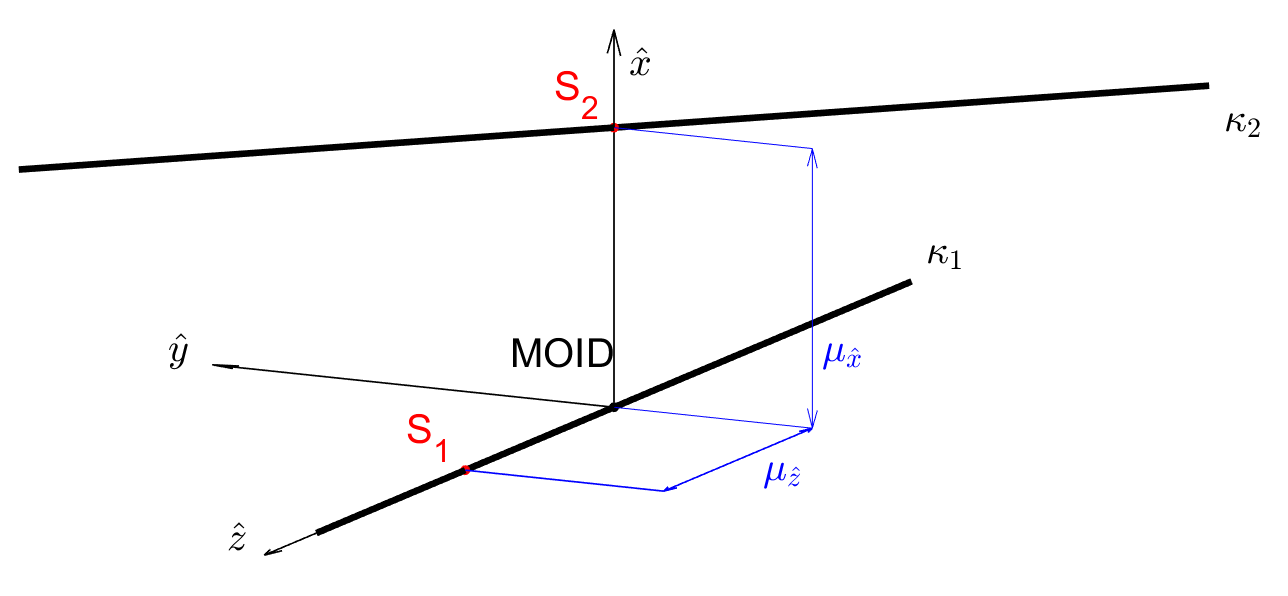}
    \caption{Miss distance components in $\Pi^{\circ}$ reference frame when $S_2$ is at the MOID.}
    \label{mu_loc}
\end{figure}

In \citet{bib_bai}, recalling Eq. (\ref{mu_x_mu_z_def}), the miss distance vector at the TCA is defined on the encounter plane with the two components $\mu_{\hat{x}}$ and $\mu_{\hat{z}}$, where:
\begin{itemize}
    \item $\mu_{\hat{x}}$ is the radial distance between $S_1$ and $S_2$
    \item $\mu_{\hat{z}}$ is the distance between $S_1$ and $S_2$ on the plane perpendicular to the radial direction
\end{itemize}
Let us consider $S_1$ to be a constellation satellite and $S_2$ a de-orbiting or injecting satellite, and let $a_1$ and $a_2$ be their respective semi-major axes. At the time instant in which $S_2$ is at the Minimum Orbit Intersection Distance \citep{bib_gronchi}, also known as MOID, depicted in Fig. \ref{mu_loc}, these two criteria are satisfied if:
\begin{subequations}
\begin{align}
    & \mu_{\hat{x}}^\otimes = \delta a
    \label{mu_moid_x}
    \\
    & \mu_{\hat{z}}^\otimes = \delta \omega a_1
    \label{mu_moid_z} \ \ \ \ \ \ \ \ \ \ \ \ \ \ \ \ \ \ \ \ \ \ \ \ \ \ \ \ \ \ \ \ \ \ \ \ \ \ \ \ \ \ \ \ \ \ \ \ \ \ \ \ \ \ \ \ \ \ \ \ \
\end{align}
\label{mu0}%
\end{subequations}
where superscript $\otimes$ denotes MOID, and:
\begin{itemize}
    \item $\delta a = a_2 - a_1$ is the radial distance between $S_1$ and $S_2$
    \item $\delta \omega = \omega_2 - \omega_1$ is the angular phase between $S_1$ and $S_2$
\end{itemize}
True anomaly phase between satellites is defined using the arguments of periapsis $\omega$, as only circular orbits are considered.
Because $S_1$ and $S_2$ move along circular orbits, only values of $\delta a$ in the order of a few kilometers typically lead to appreciable levels of collision probabilities. Therefore, $a_1$ and $a_2$ are in general very close and $S_1$ and $S_2$ can be considered to have equal velocity $v$. Moreover, for most cases, the MOID is close to the location of the NCA and the trajectories of both satellites in this interval can be considered to be linear. Under these hypotheses, the along-track distance $\mu_z(t)$ at the general time instant $t$ can be computed \citep{bib_tesi} with respect to the along-track distance $\mu_z(t_0)$ at the time instant $t_0$ at which $S_2$ is at the MOID:
\begin{equation}
\mu_{\hat{z}}^2 (t) = \mu_{\hat{z}}^2 (t_0) + (2v^2 t^2 - 2 \mu_{\hat{z}}^2 (t_0) v t)(1- \cos \varphi)
\label{mu_t}
\end{equation}
Recalling that $\mu_{\hat{x}}$ is constant, the minimum miss distance corresponds to a minimum of $\mu_{\hat{z}}^2 (t)$. Imposing the derivative of Eq. (\ref{mu_t}) equal to zero, the miss distance vector at TCA is:
\begin{subequations}
\begin{align}
    \mu_{\hat{x}} & = \delta a
    \label{mu_final_x}
    \\
    \mu_{\hat{z}} & = \delta \omega a_1 \cos \frac{\varphi}{2} \ \ \ \ \ \ \ \ \ \ \ \ \ \ \ \ \ \ \ \ \ \ \ \ \ \ \ \ \ \ \ \ \ \ \ \ \ \ \ \ \ \ \ \ \ \ \ 
    \label{mu_final_y}
\end{align}
\label{mu_final}%
\end{subequations}
As a final remark, two circular orbits always share two MOIDs which are characterised by equal and opposite nominal miss distance vectors in ECI coordinates. Moreover, these vectors lie on the same line which also intersects Earth's centre.

\subsection{Average collision probability with one constellation's satellite}
\label{mean_cp_dem}

\begin{figure}[t]
    \centering
    \includegraphics[width=.8\linewidth]{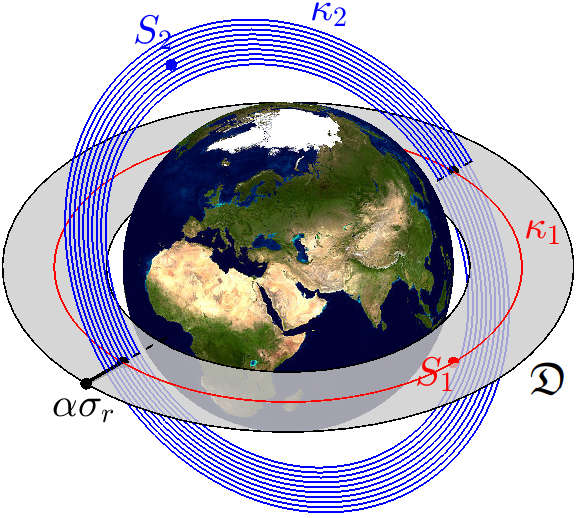}
    \caption{shell-crossing event.}
    \label{domain}
\end{figure}
The dynamics of a shell-crossing event are modeled such that the constellation satellite $S_1$ is on a fixed circular orbit, and the spiraling trajectory of the crossing satellite $S_2$, which is eventually broken down into a series of a circular orbits, would result in a series of close approaches. This is represented in Fig. \ref{domain}. For each revolution of $S_2$, a maximum of two close approaches may occur. The overall collision probability can be evaluated by considering all together the collision probabilities related to each close approach. \\
Referring to Fig. \ref{domain}, let us consider the annulus $\mathfrak{D}$ centred at $\kappa_1$ of width $2 \alpha \sigma_r$, where $\alpha$ is a parameter, to be the shell-crossing event domain. Then the mean collision probability $\overline{P}$ over the domain $\mathfrak{D}$ can be evaluated as:
\begin{equation}
    \overline{P} = \frac{1}{\mathfrak{D}} \int \iint_\mathfrak{D} P_\odot \exp \left( -\frac{1}{2} \frac{\mu_{\hat{x}}^2}{\sigma_{\hat{x}}^2} \right) \exp \left( -\frac{1}{2} \frac{\mu_{\hat{z}}^2}{\sigma_{\hat{z}}^2} \right) \textrm{d}\mu_{\hat{x}} \textrm{d}\mu_{\hat{z}}
    \label{p_integral}
\end{equation}
Substituting Eq.s (\ref{mu_final_x}) and (\ref{mu_final_y}) into Eq. (\ref{p_integral}) and introducing the new variable $\delta z = a_1 \delta \omega$ yields:
\begin{equation}
    \overline{P} = \frac{P_\odot}{\mathfrak{D}} \int_\mathfrak{D} \exp \left( -\frac{1}{2} \frac{\delta a^2}{\sigma_r^2} \right) \exp \left( -\frac{1}{2} \frac{\delta z^2}{\sigma_{\theta}^2} \right) \textrm{d}(\delta a) \textrm{d}(\delta z)
    \label{p_integral_final}
\end{equation}
where:
\begin{subequations}
\begin{align}
    \sigma_r^2 & = \sigma_{\hat{x}}^2 = \sigma_{1_R}^2 + \sigma_{2_R}^2
    \label{sigma_r_def}
    \\
    \sigma_\theta^2 & = \frac{\sigma_{\hat{z}}^2}{\cos^2 \frac{\varphi}{2}} = (\sigma_{1_S}^2 + \sigma_{2_S}^2) + (\sigma_{1_W}^2 + \sigma_{2_W}^2) \tan^2 \frac{\varphi}{2}
    \label{sigma_theta_def}
\end{align}
\label{sigma_2}%
\end{subequations}
The area of the annulus $\mathfrak{D}$ is:
\begin{equation}
    \mathfrak{D} = \pi (a_1 + \alpha \sigma_r)^2 - \pi (a_1 - \alpha \sigma_r)^2 = 4 \pi a_1 \alpha \sigma_r
    \label{area_d}
\end{equation}
Therefore, Eq. (\ref{p_integral_final}) can be written as:
\begin{align}
\begin{split}
    \overline{P} & = \frac{P_\odot}{4 \pi a_1 \alpha \sigma_r} \int_{-\alpha \sigma_r}^{\alpha \sigma_r} \exp \left( -\frac{1}{2} \frac{\delta a^2}{\sigma_r^2} \right) \textrm{d}(\delta a) \int_{-\pi a_1}^{\pi a_1} \exp \left( -\frac{1}{2} \frac{\delta z^2}{\sigma_\theta^2} \right) \textrm{d}(\delta z)
    \\
    & = \frac{1}{\alpha}\frac{P_\odot}{2} \frac{\sigma_\theta}{a_1} \textrm{erf} \left( \frac{1}{\sqrt{2}}\alpha \right) \textrm{erf} \left( \frac{1}{\sqrt{2}} \frac{\pi a_1}{\sigma_\theta} \right)
    \\
    & \approx \frac{1}{\alpha}\frac{P_\odot}{2} \frac{\sigma_\theta}{a_1} \textrm{erf} \left( \frac{1}{\sqrt{2}}\alpha \right)
    \label{p_mean}
\end{split}
\end{align}
For typical satellite applications $a \gg \sigma_\theta$ and $\textrm{erf} \left( \frac{1}{\sqrt{2}} \frac{\pi a_1}{\sigma_\theta} \right) \approx 1$. \\
Following the same reasoning, a close approach will occur if both $S_1$ and $S_2$ are within $\mathfrak{D}$ at the same instant. As $S_1$ always lies on $\mathfrak{D}$, a close approach will occur when $S_2$ crosses the domain $\mathfrak{D}$, that is when $S_2$ is at the MOID. It follows that for every revolution of $S_2$, two close approaches will occur and:
\begin{equation}
    \overline{N}_{CA} = 2 R_{\mathfrak{D}}
    \label{rd_def}
\end{equation}
where $R_{\mathfrak{D}}$ is the number of revolutions of $S_2$ that intersect the domain $\mathfrak{D}$. \\
The number of revolutions $R_{\mathfrak{D}}$ is computed as the ratio between the width of the annulus $\mathfrak{D}$ and the semi-major axis variation per revolution $|\Delta a|$ of $S_2$,therefore Eq. (\ref{rd_def}) can be written as:
\begin{equation}
    \overline{N}_{CA} = 4 \frac{\alpha \sigma_r}{|\Delta a|}
    \label{nca_linear_nca}
\end{equation}
and the semi-major axis variation, which is assumed to be constant during the shell-crossing event, is:
\begin{equation}
    \Delta a = \dot a T_1
    \label{delta_a_T}
\end{equation}
where $T_1$ is the orbital period of $S_1$. \\
Finally, the mean collision probability $P_{satellite}$ between a crossing satellite and a constellation satellite can be obtained by imposing the domain $\mathfrak{D}$ to have infinite radial extension, thus:
\begin{align}
\begin{split}
    P_{satellite} & = \lim_{\alpha \to + \infty} \left\{ 1 - \left[1 - \frac{1}{\alpha}\frac{P_\odot}{2} \frac{\sigma_\theta}{a_1} \textrm{erf} \left( \frac{\alpha}{\sqrt{2}} \right)\right]^{4 \frac{\alpha \sigma_r}{|\Delta a|}} \right\}
    \\
    & = 1 - \exp \left( - 2 \frac{P_\odot \sigma_r \sigma_\theta}{|\Delta a| a_1} \right)
    \end{split}
    \label{p_plane1}
\end{align}

\subsection{Head-on collisions}
\label{hocc}

\begin{figure}[t]
    \centering
    \includegraphics[width=.8\linewidth]{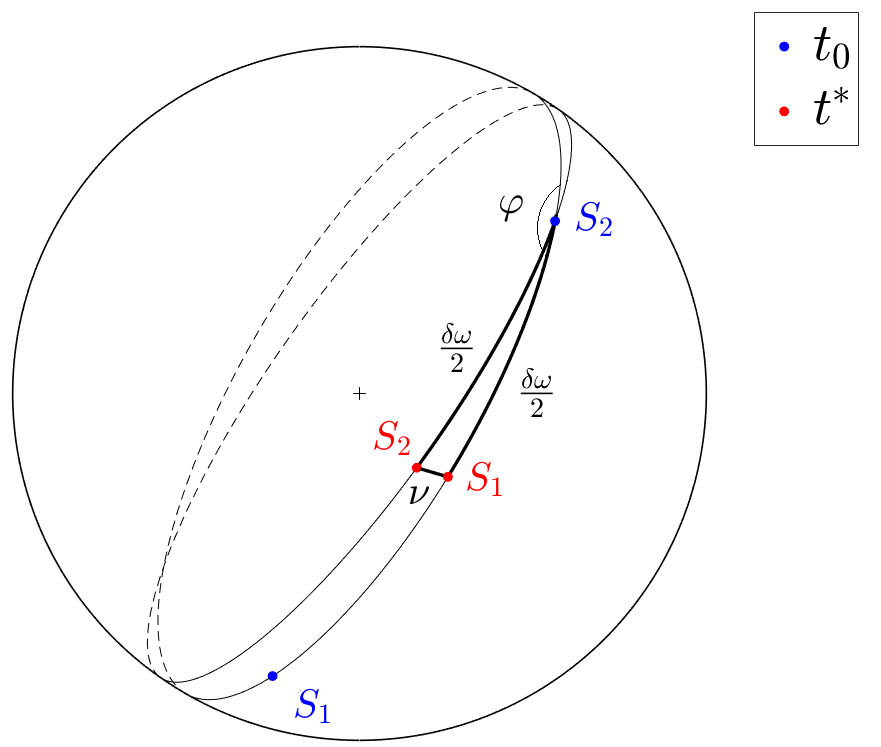}
    \caption{Non-scaled representation of NCA for near straight collision angles $\varphi$.}
    \label{spherical}
\end{figure}
The hypothesis of linear motion of both satellites between MOID and NCA in Section \ref{mean_cp_dem}, cannot be considered valid anymore in case of head-on collisions, as the collision PDF is not condensed in the vicinity of the MOID. That is when $\kappa_1$ and $\kappa_2$ are characterised by (almost) parallel angular momenta, but with opposite directions. \\
The time instants $t_0$ and $t^*$ in Fig. \ref{spherical}, respectively refer to the time instant in which $S_2$ is at the MOID and the TCA. As before, assuming both the constellation's and crossing satellites to have equal angular velocity, the angle $\nu(t)$ represented in Fig. \ref{spherical} can be expressed using the cosine rule as:
\begin{equation}
    \cos \nu(t) = \cos (nt) \cos(\delta\omega - nt) + \sin(nt) \sin(\delta \omega -nt)\cos(\pi - \varphi)
    \label{cosnu_g}
\end{equation}
Imposing the derivative of Eq. (\ref{cosnu_g}) equal to zero, as previously done in Section \ref{NCA_sec}, and recalling that for head-on collisions $\mu_{\hat{z}} = a_1 \nu$, yields:
\begin{equation}
    \mu_{\hat{z}} = 2 a_1 \sin \frac{\delta \omega}{2} \cos \frac{\varphi}{2}
    \label{nca_sph}
\end{equation}
Finally, using the same method from Section \ref{mean_cp_dem} for the evaluation of $\overline{P}$ and, again, considering the annular domain $\mathfrak{D}$ to have infinite radial extension, the overall collision probability $P_{satellite}$ between one constellation's and one crossing satellite can be evaluated as:
\begin{equation}
    P_{satellite} = 1 - \exp \left[ -2 \sqrt{2 \pi} \frac{P_\odot \sigma_r}{|\Delta a|}\exp\left(-\frac{a_1^2}{\sigma_\theta^2}\right) I_0 \left( \frac{a_1^2}{\sigma_\theta^2} \right) \right]
    \label{p_plane2}
\end{equation}
where $I_0$ is the modified Bessel function of the first kind of order zero. \\

\subsection{Average collision probability of a shell-crossing event}

\begin{figure}[t]
    \centering
    \includegraphics[width=.8\linewidth]{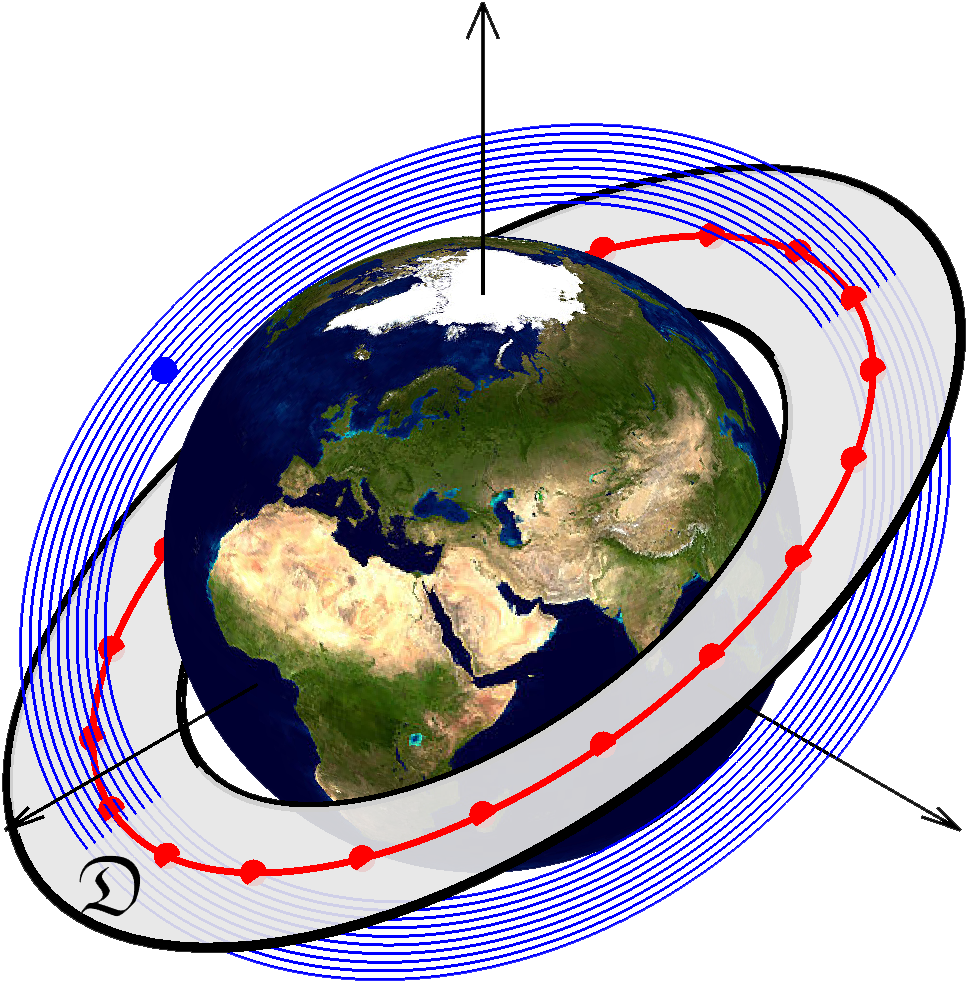}
    \caption{One constellation's orbital plane with $N_S$ satellites.}
    \label{orbital_plane}
\end{figure}
In general, more than one satellite moves along the orbit $\kappa_1$ of a satellite constellation  \citep{bib_walker}. Let $N_S$ be the number of satellites on one constellation's orbital plane, as shown in Fig. \ref{orbital_plane}. \\
The average collision probability $P_{plane}$ with one constellation's orbital plane can be evaluated by using Eq.s (\ref{p_plane1}) and (\ref{p_plane2}) for each one of the $N_S$ satellites. Nevertheless, the evaluation is simplified if all the $N_S$ satellites are assumed to have equal error ellipsoid in their respective RSW reference frames. Indeed, under this hypothesis, the crossing satellite has the same average collision probability with each one of the $N_S$ constellation's satellites belonging to a single orbital plane. Assuming the collision probabilities with each satellite to be independent from one another, then:
\begin{equation}
    P_{plane} = 1 - \left( 1 - P_{satellite} \right)^{N_S}
\end{equation}
Referring to the general case described in Section \ref{mean_cp_dem}, $P_{plane}$ can be computed as:
\begin{equation}
    P_{plane} = 1 - \exp \left( -2 \frac{P_\odot N_S \sigma_r \sigma_\theta}{|\Delta a| a_1} \right)
    \label{P_op_l}
\end{equation}
whereas for the head-on collision case from Section \ref{hocc}:
\begin{equation}
    P_{plane} = 1 - \exp \left[ -2 \sqrt{2 \pi} \frac{P_\odot N_S \sigma_r}{|\Delta a|}\exp\left(-\frac{a_1^2}{\sigma_\theta^2}\right) I_0 \left( \frac{a_1^2}{\sigma_\theta^2} \right) \right]
    \label{P_op_s}
\end{equation}
Finally, the average collision probability $P_{shell}$ related to a shell-crossing event can be computed as:
\begin{equation}
    P_{shell} = {\displaystyle 1 - \prod_{i=1}^{N_P} \left( 1 - P_{plane,i} \right)}
    \label{p_final}
\end{equation}
where $N_P$ is the total number of orbital planes of the constellation. \\

\subsection{Threshold collision angle}

The expression of $P_{plane,i}$ in Eq. (\ref{p_final}) depends on the angle $\varphi_i$ between the crossing orbital plane and the $i$-th orbital plane of the constellation. It must be emphasised that Eq. (\ref{P_op_l}) holds if $\varphi \napprox \pi$, whereas Eq. (\ref{P_op_s}) is always valid. Nonetheless, the expressions $\exp\left(-\frac{a_1^2}{\sigma_\theta^2}\right)$ and $I_0 \left( \frac{a_1^2}{\sigma_\theta^2} \right)$ exponentially decrease and increase, respectively, the smaller $\varphi$ is. This would result in numerical values out of the floating-point range normally supported by most computational software. \\
It follows that a threshold value $\varphi^*$ has to be defined to distinguish whether Eq. (\ref{P_op_l}) or Eq. (\ref{P_op_s}) has to be used.
In this study, MATLAB is used for constellation risk assessments, which has a largest positive floating-point number of around $1.7\times10^{308}$ (double precision). Let:
\begin{equation}
\Phi_{max} = \frac{a_1}{\sigma_\theta (\varphi^*)}
\end{equation}
such that $I_0 (\Phi^2_{max}) < N_{fp,\text{max}}$, where $N_{fp,\text{max}}$ is the largest floating-point number available. \\
Recalling equation\ref{sigma_theta_def}, the threshold angle $\varphi^*$ can be computed as:
\begin{equation}
    \varphi^* = 2 \tan^{-1} \sqrt{\frac{1}{\sigma_W^2}\left(\frac{a_1^2}{\Phi_{max}^2} - \sigma_S^2 \right)}
    \label{phi_crit_mat_final}
\end{equation}
where:
\begin{subequations}
\begin{align}
    \sigma_S^2 & = \sigma_{1_S}^2 + \sigma_{2_S}^2
    \\
    \sigma_W^2 & = \sigma_{1_W}^2 + \sigma_{2_W}^2
\end{align}
\label{S_W_sub}%
\end{subequations}
In the following sections, a value of $\Phi_{max} = 12.5$ has been used. \\
An alternative strategy involves the definition of a maximum relative error $\varepsilon_\Phi$ between Eq.s (\ref{P_op_l}) and (\ref{P_op_s}). The threshold angle $\varphi^*$ can be then computed as:
\begin{equation}
    \tan^2 \frac{\varphi^*}{2} \approx 8\frac{a_1^2}{\sigma_{W}^2} \frac{\varepsilon_\Phi}{1 - \varepsilon_\Phi}
    \label{phicrit_error}
\end{equation}
The continuity between Eq.s (\ref{P_op_l}) and (\ref{P_op_s}) in a neighborhood of $\varphi^*$ can also be proven, as shown in \citet{bib_tesi}. \\
Moreover, if $\varphi_i \leq \varphi^*$ for all $i$, which means the absence of potential heads-on collisions, then, substituting Eq. (\ref{P_op_l}) into Eq. (\ref{p_final}), and considering its Maclaurin expansion:
\begin{equation}
    P_{shell} \approx \frac{N_S r_a^2}{|\Delta a| a_1} {\displaystyle \sum_{i=1}^{N_P} \frac{1}{\cos \frac{\varphi_i}{2}}}
    \label{final3_eq}
\end{equation}
where $P_{shell}$ does not depend anymore on the position uncertainties of the crossing and constellation's satellites.

\subsection{Summary of hypotheses}
\label{hype}

The model developed in this section is based on the following assumptions:
\begin{enumerate}
    \item All the effects of natural perturbations are neglected, with the only exception of altitude variation due to air drag
    \item All satellites are assumed to move along circular orbits
    \item The altitude drop per orbit of $S_2$ inside $\mathfrak{D}$ is considered to be constant
    \item The covariance matrices of the position error of both satellites are assumed to be diagonal in their respective RSW reference frames \citep{bib_chan}
    \item The satellites are modeled as spheres
    \item The collision probabilities of each close approach are assumed to be independent from one another
\end{enumerate}

\section{MOID propagation model}
\label{moidprop_sec}

\begin{figure}[t]
    \centering
    \includegraphics[width=\linewidth]{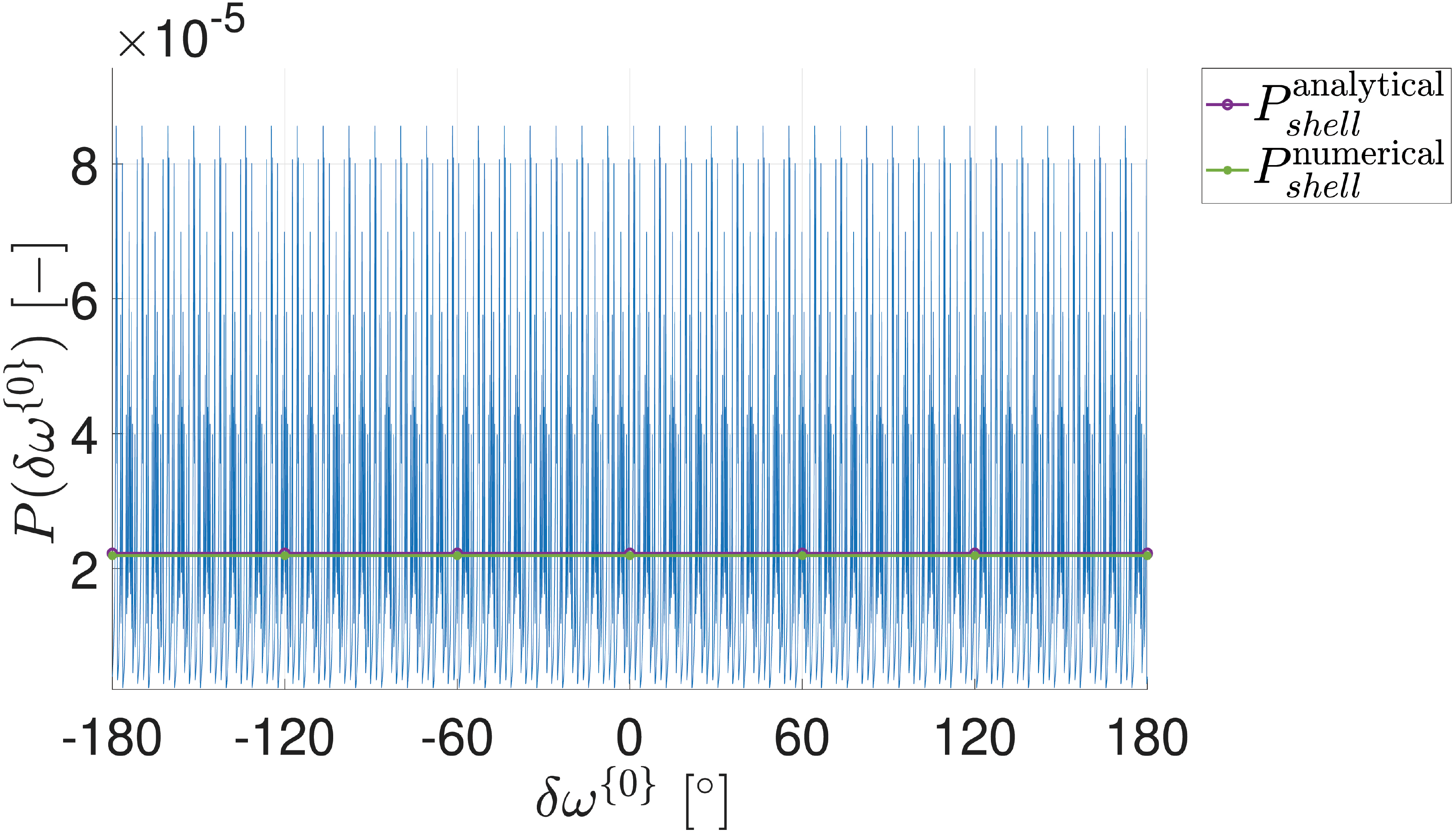}
    \caption{Collision probability distribution $P(\delta \omega^{\{0\}})$ and its mean value $P_{shell}$ of a satellite crossing OneWeb constellation with $|\Delta a| = 0.1$ km/s.}
    \label{validation}
\end{figure}
The statistical model developed in Section \ref{model_sec} is an efficient tool for risk assessments of highly populated traffic scenarios. On the other hand, it fails to consider phase between satellites. For this purpose, the MOID propagation model has been developed. \\
This is a simplified propagation model whose time step is equal to half the orbital period of the constellation satellite. Each iteration is composed of three main steps:
\begin{itemize}
    \item altitude variation of $S_1$
    \item phase shift between $S_1$ and $S_2$ due to different orbital periods
    \item collision probability assessment at NCA
\end{itemize}
Significant values of collision probabilities require both objects to be relatively close to one another. Therefore, considering the orbits to be circular within the propagation time step, it follows that $a_1 \approx a_2$. With this consideration in mind, the semi-major axis of $S_2$ at each time step would be:
\begin{equation}
    a_2^{\{k+1\}} = a_2^{\{k\}} + \frac{\Delta a}{2}
\end{equation}
Following the same reasoning, $S_1$ true anomaly can be estimated as:
\begin{equation}
    \theta^{\{k+1\}} = \theta^{\{k\}} + \pi \left( \frac{a_1}{a_2^{\{k\}}} \right)^{\tfrac{3}{2}}
\end{equation}
Let the constellation satellite be at one of the two MOIDs at the initial propagation time $t=0$, then at each step it will always be located, alternately, at the two MOIDs. Under this condition, the miss distance of every NCA can be computed with Eq.s (\ref{mu_final_x}) and (\ref{mu_final_y}), or Eq. (\ref{nca_sph}) if $\varphi>\varphi*$, where:
\begin{subequations}
\begin{align}
    \delta a^{\{k+1\}} & = \delta a^{\{k\}} + \frac{\Delta a}{2} \\
    \delta \omega^{\{k+1\}} & = \delta \omega^{\{k\}} + \pi \left[ 1 - \left( \frac{a_1}{a_2^{\{k\}}} \right)^{\tfrac{3}{2}} \right]  \ \ \ \ \ \ \ \ \ \ \ \ \ \ \ \  \ \ \ \ \ \ \ \ 
\end{align}%
\end{subequations}
The overall collision probability $P(\delta \omega^{\{0\}})$ can be evaluated through the combination of the collision probabilities of each NCA, computed with Chan's algorithm truncated at the $4^{\text{th}}$ term. Repeating the process for different initial true anomaly phases $\delta\omega^{\{0\}}$, the collision probability profile shown in Fig. \ref{validation} is obtained. The analytical evaluation of $P_{shell}$ from Fig. (\ref{validation}) refers to Eq. (\ref{p_final}), whereas its numerical value is given as the integral of the distribution obtained with the MOID propagation model. It has been extensively demonstrated that these two always correspond \citep{bib_tesi}, hence, the following relation holds:
\begin{equation}
    P_{shell} = {\displaystyle 1 - \prod_{i=1}^{N_P} \left( 1 - P_{plane,i} \right)} \approx \frac{1}{2 \pi} \int_0^{2 \pi} P(\delta \omega^{\{0\}}) \textrm{d}(\delta \omega^{\{0\}})
    \label{validation_eq}
\end{equation}
Eq. (\ref{validation_eq}) has been validated for different constellations, $|\Delta a|$, $\sigma_1^{\textrm{RSW}}$, $\sigma_2^{\textrm{RSW}}$ and orientations of the crossing orbital plane. \\
Nonetheless, it must be emphasised that Eq. (\ref{validation_eq}) cannot be considered always valid, as large values of semi-major axis variation per revolution $|\Delta a|$, would result in a fewer number of close approaches. In this context, a statistical approach fails to be consistent within acceptable error boundaries, as the mean collision probability over the true anomaly phase also depends on the initial radial distance $\delta a^{\{0\}}$. \\
It has been observed that, recalling Eq. (\ref{sigma_r_def}), Eq. (\ref{validation_eq}) holds with a relative error lower than $1\times10^{-3}$, if:
\begin{equation}
    \frac{3\sigma_r}{|\Delta a|} \geq 1
    \label{sma_condition}
\end{equation}
In \citet{bib_tesi} it is shown that, if Eq. (\ref{sma_condition}) is not satisfied, the average collision probability over the true anomaly phase follows a sinusoidal-like behaviour which depends on $\delta a^{\{0\}}$, and has a period equal to $|\Delta a|$. Under this condition, the statistical model presented in Section \ref{model_sec} assesses the mean collision probability, which is averaged over both $\delta \omega^{\{0\}}$ and $\delta a^{\{0\}}$. \\
A series of MC simulations using different values of $\mu_{\hat{x}}$ and $\mu_{\hat{z}}$ have also been performed in order to validate the assumption of using Chan's algorithm truncated at the first term \citep{bib_tesi}. \\
From Fig. \ref{validation} is clear how the true anomaly phase $\delta \omega$ plays a key role for collision probability minimisation. The MOID propagation model also constitutes an efficient tool for collision probability assessments, as the collision probability spectrum over the true anomaly phase can be outlined with extremely low computational effort. \\
Indeed, the larger the semi-major axis rate of change $|\Delta a|$, the more efficient the MOID propagation model, since fewer close approaches have to  be assessed. Conveniently, that is the case in which a statistical approach loses its validity.

\section{Validation}
\label{valid_sec}

\begin{figure}
    \centering
    \includegraphics[width=\linewidth]{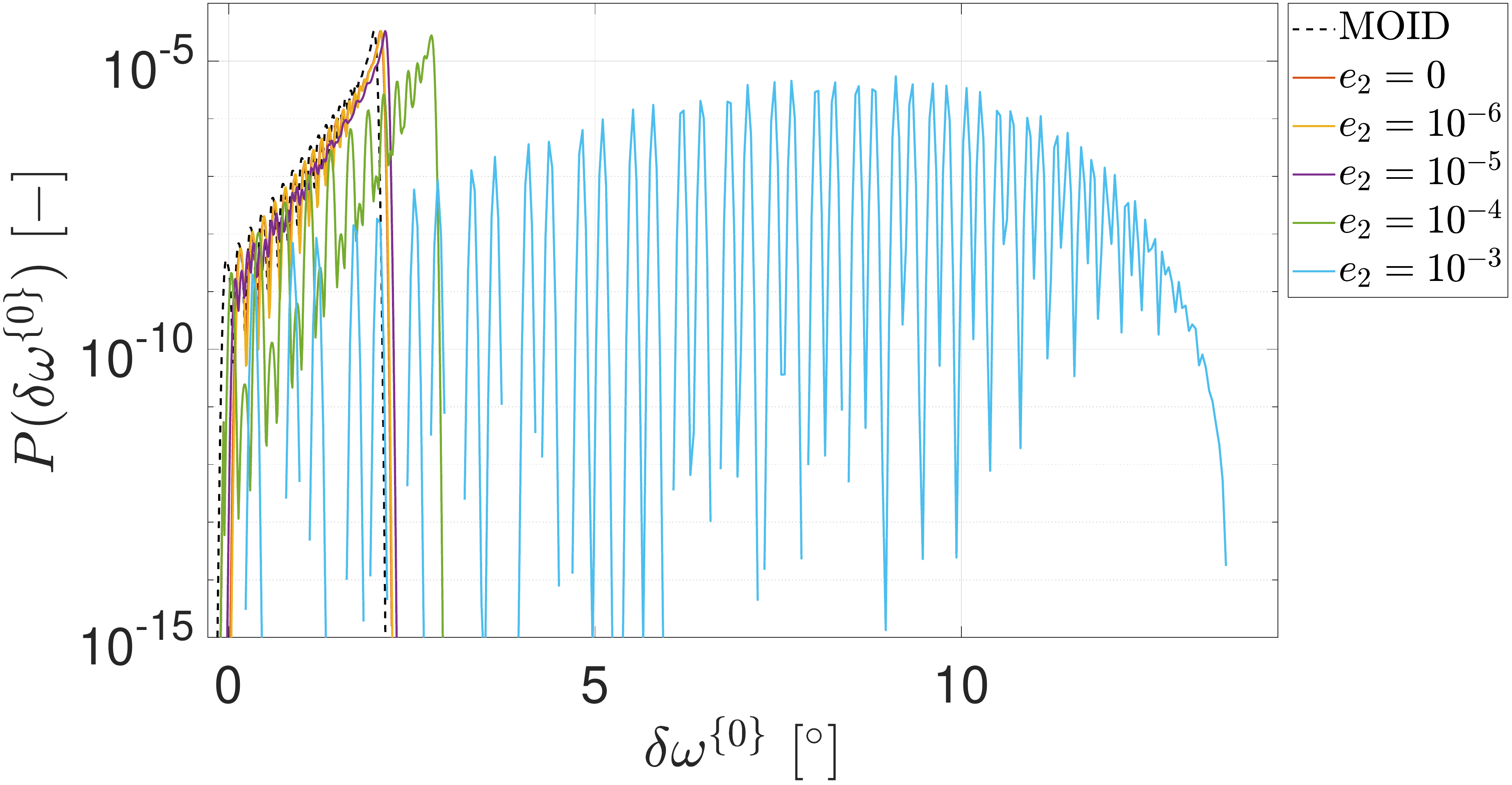}
    \caption{Comparison between the collision probability profiles $P(\delta \omega^{\{0\}})$ obtained with the MOID propagation model, and standard propagation assuming different initial eccentricity of the crossing satellite. Simulation data: $\varphi = \tfrac{\pi}{6}$, $\boldsymbol{\zeta}_1^{\textrm{RSW}}=\textrm{diag}\{\sfrac{1}{4},1,\sfrac{1}{4}\}$ km$^2$, $\boldsymbol{\zeta}_2^{\textrm{RSW}}=\textrm{diag}\{1,4,1\}$ km$^2$, $r_a = 4$, $C_D = 2.2$, $P_t = 400$ W, $\eta = 0.5$, $I_{sp}=3000$ s.}
    \label{p_omega}
\end{figure}

Most of the assumptions on which the statistical model is based, listed in Section \ref{hype}, do not considerably change the dynamics of the shell-crossing event. The most stringent hypotheses are the first two: neglecting other natural sources of perturbation ($J_2$ in particular) and discretising the spiraling trajectory of the crossing satellite into circular orbits. In this section, the validity of the latter is assessed, for different values of eccentricity of the crossing orbit. \\
The most suitable option for validation would be a series of MC simulations for different values of initial phase $\delta \omega^{\{0\}}$. Then, the average collision probability could be computed with Eq. (\ref{validation_eq}). Nevertheless, billions of MC runs would be necessary to analyse the collision probability profile over the initial phase $\delta \omega^{\{0\}}$, and this process would have to be repeated for different relative orientations between $\kappa_1$ and $\kappa_2$, i. e. for different values of $\varphi$. \\
As this would require to much computational power, a different approach is hereby used. The motion of two satellites, one from the constellation and one crossing it, are propagated through the integration of the equations of motion in Cartesian coordinates. Once the nominal orbits are known, NCAs are detected and their respective collision probability is computed with Chan's algorithm truncated at the $4^{\textrm{th}}$ term. The position uncertainties of the satellites are considered to be constant in their respective RSW frame, during the duration of the shell-crossing event. \\
It is worth mentioning that the error introduced by this last assumption may affects the results locally, but the overall effect on the average collision probability could be negligible. Further investigations about this topic are desirable, but limited by the great computational effort required. \\
In Fig. \ref{p_omega} is represented the collision probability spectrum $P(\delta \omega^{\{0\}})$ evaluated with both MOID and standard propagation models. The latter has been used for different values of initial orbit eccentricity $e_2$. Simulations were carried out considering two Starlink satellites of mass $386$ kg and cross-sectional radius of $2.39$ m \citep{bib_lemay}, both belonging to the Starlink (4) shell from Table \ref{catalogue}. The crossing satellite is de-orbiting. \\
The collision probability profiles $P(\delta \omega^{\{0\}})$ were used to evaluate the average collision probabilities of the shell-crossing events, exploiting trapezoidal numerical integration. These results are listed in Table \ref{table_validation}, from which it can be noticed that the statistical model is able to predict with great accuracy the collision probability for crossing orbit eccentricity up to $10^{-3}$. It was not possible to test larger values of eccentricity due to the considerable computational effort required. Moreover, the larger $e_2$, the less accurate is the result, as a wider interval of initial true anomaly phases $\delta\omega^{\{0\}}$ decreases the resolution of the collision probability profile $P(\delta\omega^{\{0\}})$. \\
Considering Fig. \ref{p_omega}, a larger eccentricity $e_2$ affects the collision probability spectrum by spreading lower values of $P(\delta \omega^{\{0\}})$ over a wider range of $\delta \omega^{\{0\}}$, keeping its mean value constant. Hence, when crossing crowded constellations, the lower $e_2$, the more the collision probability can be minimised through the proper selection of $\delta\omega^{\{0\}}$. \\
A final remark has to be made about gravitational perturbations due to Earth's oblateness, as it strongly affects the dynamics of low-altitude orbits. It was observed that the altitude variation caused by considering $J_2$, decreases the probability per revolution that a close approach will occur, but increases the number of overall revolutions in the vicinity of the constellation's orbit $\kappa_1$. Further analyses may be the subject for future studies.

\begin{table*}
\centering
\caption{Catalogue of the satellite constellations considered in the study.}
\begin{tabular}{|l|l|l|l|l|l|l|l|c|}
\hline

Constellation & $i_1 \ [^{\circ}]$ & $N_T$ & $N_P$ & $f$ & $h \ [\textrm{km}]$ & Shell & Status$^{\text{a}}$ & Reference
\\ \hline

\multirow{8}*{Starlink} & 42 & 2493 & 42 & 2 & 336 & (1) & A & \multirow{8}*{\citet{bib_starlink}} \\ \cline{2-8}
& 48 & 2478 & 42 & 2 & 341 & (2) & A &  \\ \cline{2-8}
& 53 & 2547 & 42 & 2 & 346 & (3) & A &
\\ \cline{2-8}
& 53.2 & 1584 & 72 & 2 & 540 & (4) & A & \\ \cline{2-8}
& 53 & 1584 & 72 & 2 & 550  & (5) & D & \\ \cline{2-8}
& 97.6 & 348 & 6 & 2 & 560  & (6) & A & \\ \cline{2-8}
& 97.6 & 172 & 4 & 2 & 565  & (7) & A & \\ \cline{2-8}
& 70 & 720 & 36 & 2 & 570 & (8) & A &
\\ \hline

\multirow{3}*{Kuiper} & 33 & 784 & 28 & 2 & 590 & (1) & A & \multirow{3}*{\citet{bib_kuiper}} \\ \cline{2-8}
& 42 & 1296 & 36 & 2 & 610 & (2) & A & \\ \cline{2-8}
& 51.9 & 1156 & 34 & 2 & 630 & (3) & A &
\\ \hline

\multirow{2}*{Telesat} & 98.98 & 351 & 27 & 3 & 1015 & (1) & C & \multirow{2}*{\citet{bib_telesat}} \\ \cline{2-8}
& 50.88 & 1320 & 33 & 3 & 1320  & (2) & A &
\\ \hline

OneWeb & 87.9 & 720 & 18 & 2 & 1200 & - & D & \citet{bib_oneweb}
\\ \hline

Kepler & 89.5 & 360 & 12 & 2 & 600 & - & A & \citet{}
\\ \hline

Iridium NEXT & 86.4 & 66 & 6 & 2 & 770 & - & C & \citet{bib_iridium}
\\ \hline

Globalstar & 52 & 48 & 8 & 2 & 1414 & - & C & \citet{bib_globalstar}
\\ \hline

\multirow{5}*{OrbcommG1} & 45 & 12 & 3 & 3 & 775 & (1) & C  & \multirow{5}*{\citet{bib_orbcomm}} \\ \cline{2-8}
& 108 & 2 & 1 & 1 & 780 & (2) & C & \\ \cline{2-8}
& 70 & 2 & 1 & 1 & 785 & (3) & C & \\ \cline{2-8}
& 45 & 24 & 3 & 3 & 820 & (4) & C & \\ \cline{2-8}
& 0 & 8 & 1 & 1 & 825 & (5) & C &
\\ \hline

Capella Space & 98 & 36 & 12 & 2 & 495 & - & D & \citet{bib_capella}
\\ \hline

\multirow{4}*{Swarm Tech} & 45 & 20 & 1 & 1 & 450 & (1) & C  & \multirow{4}*{\citet{bib_swarmtech}} \\ \cline{2-8}
& 10 & 20 & 1 & 1 & 500 & (2) & C & \\ \cline{2-8}
& 97.4 & 62 & 1 & 1 & 505 & (3) & C & \\ \cline{2-8}
& 97.6 & 48 & 1 & 1 & 555 & (4) & C &
\\ \hline

\multirow{3}*{Planet} & 51.6 & 28 & 1 & 1 & 410 & (1) & C  & \multirow{3}*{\citet{bib_planet}} \\ \cline{2-8}
& 51.6 & 28 & 1 & 1 & 415 & (2) & C & \\ \cline{2-8}
& 97.98 & 11 & 1 & 1 & 620 & (3) & C &
\\ \hline

\multirow{3}*{HawkEye 360} & 14.25 & 2 & 1 & 1 & 575 & (1) & C  & \multirow{3}*{\citet{bib_hawkeye}} \\ \cline{2-8}
& 45 & 10 & 5 & 5 & 580 & (2) & C & \\ \cline{2-8}
& 14.25 & 2 & 1 & 1 & 585 & (3) & C &
\\ \hline

\end{tabular}
\\
\medskip
\footnotesize
$^\text{a}$(A) Approved (D) Deployment (C) Completed
\label{catalogue}
\end{table*}

\begin{table*}
\centering
\caption{Shell-crossing event average collision probability evaluated with Cartesian propagation and Chan's algorithm for different initial crossing eccentricity $e_2$ and collision angle $\varphi$.
Respective values of $P_{shell}$ are reported for comparison.}
\begin{tabular}{|l|l|l|l|l|l|l|}
\hline

\diagbox[width=7em]{$e_2$ [-]}{$\varphi$ [rad]} & $\tfrac{\pi}{6}$ & $\tfrac{\pi}{3}$ & $\tfrac{\pi}{2}$ & $\tfrac{2\pi}{3}$ & $\tfrac{5\pi}{6}$ & $\pi$ \\ \hline

$0$ & $0.91218\times10^{-8}$ &  $0.10177\times10^{-7}$ & $0.12469\times10^{-7}$ & $0.17638\times10^{-7}$ & $0.34075\times10^{-7}$ & $0.13678\times10^{-3}$ \\ \hline

$10^{-6}$ & $0.91246\times10^{-8}$ &  $0.10179\times10^{-7}$ & $0.12470\times10^{-7}$ & $0.17638\times10^{-7}$ & $0.34075\times10^{-7}$ & $0.13678\times10^{-3}$ \\ \hline

$10^{-5}$ & $0.91423\times10^{-8}$ &  $0.11019\times10^{-7}$ & $0.12477\times10^{-7}$ & $0.17639\times10^{-7}$ & $0.34076\times10^{-7}$ & $0.13679\times10^{-3}$ \\ \hline

$10^{-4}$ & $0.91404\times10^{-8}$ &  $0.10192\times10^{-7}$ & $0.12477\times10^{-7}$ & $0.17640\times10^{-7}$ & $0.34078\times10^{-7}$ & $0.13678\times10^{-3}$ \\ \hline

$10^{-3}$ & $0.90144\times10^{-8}$ &  $0.10077\times10^{-7}$ & $0.12393\times10^{-7}$ & $0.17614\times10^{-7}$ & $0.34076\times10^{-7}$ & $0.13673\times10^{-3}$ \\ \hline \hline

\textbf{$P_{shell}$} & $0.91313\times10^{-8}$ &  $0.10185\times10^{-7}$ & $0.12474\times10^{-7}$ & $0.17640\times10^{-7}$ & $0.34078\times10^{-7}$ & $0.13680\times10^{-3}$ \\ \hline

\end{tabular}
\label{table_validation}
\end{table*}

\begin{figure}[b!]
    \centering
    \includegraphics[width=\linewidth]{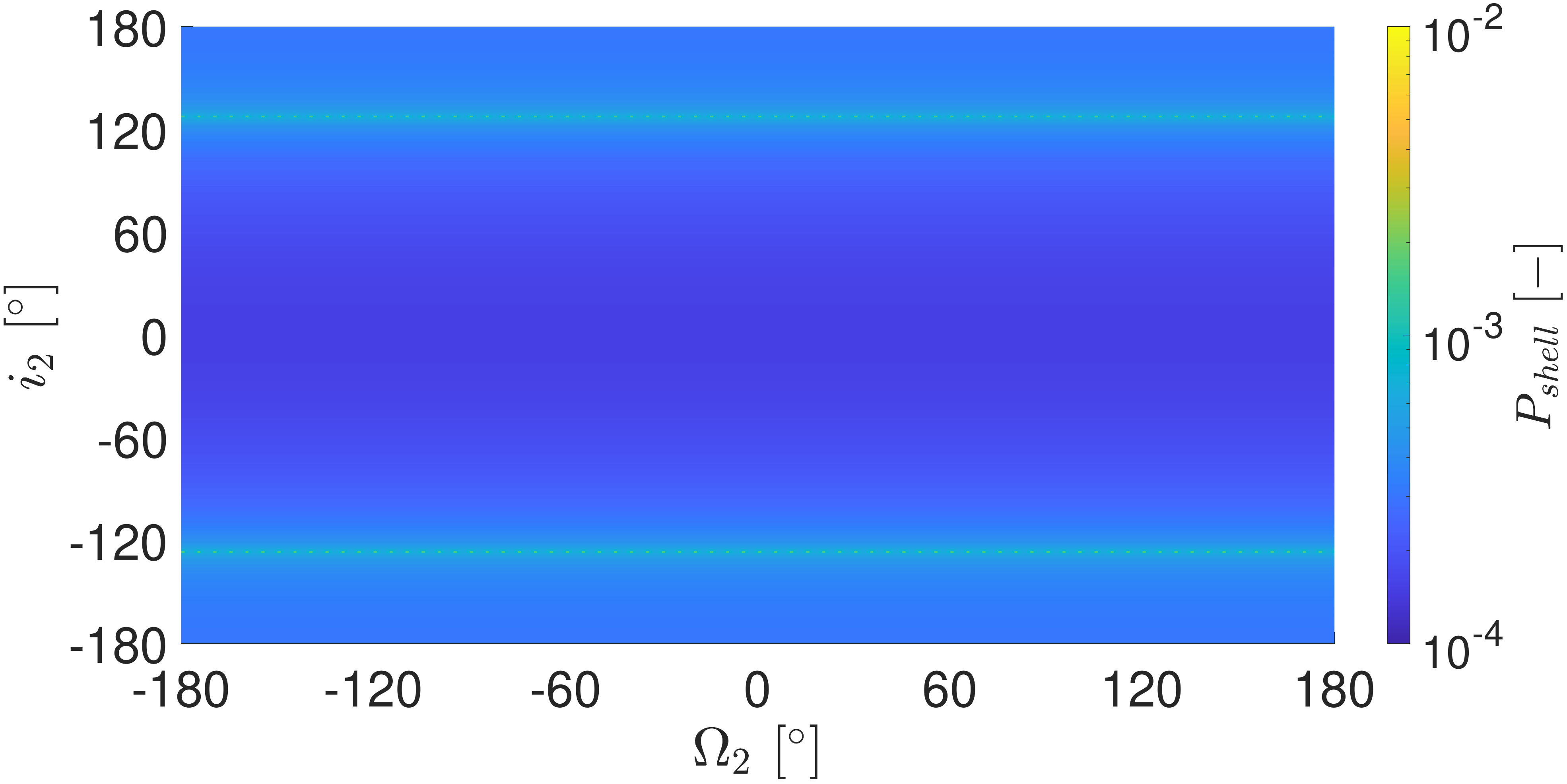}
    \caption{Collision probability distribution of passive de-orbiting through Starlink (4) shell.}
    \label{s4_cross_distrib}
\end{figure}
\begin{figure}[b!]
    \centering
    \includegraphics[width=\linewidth]{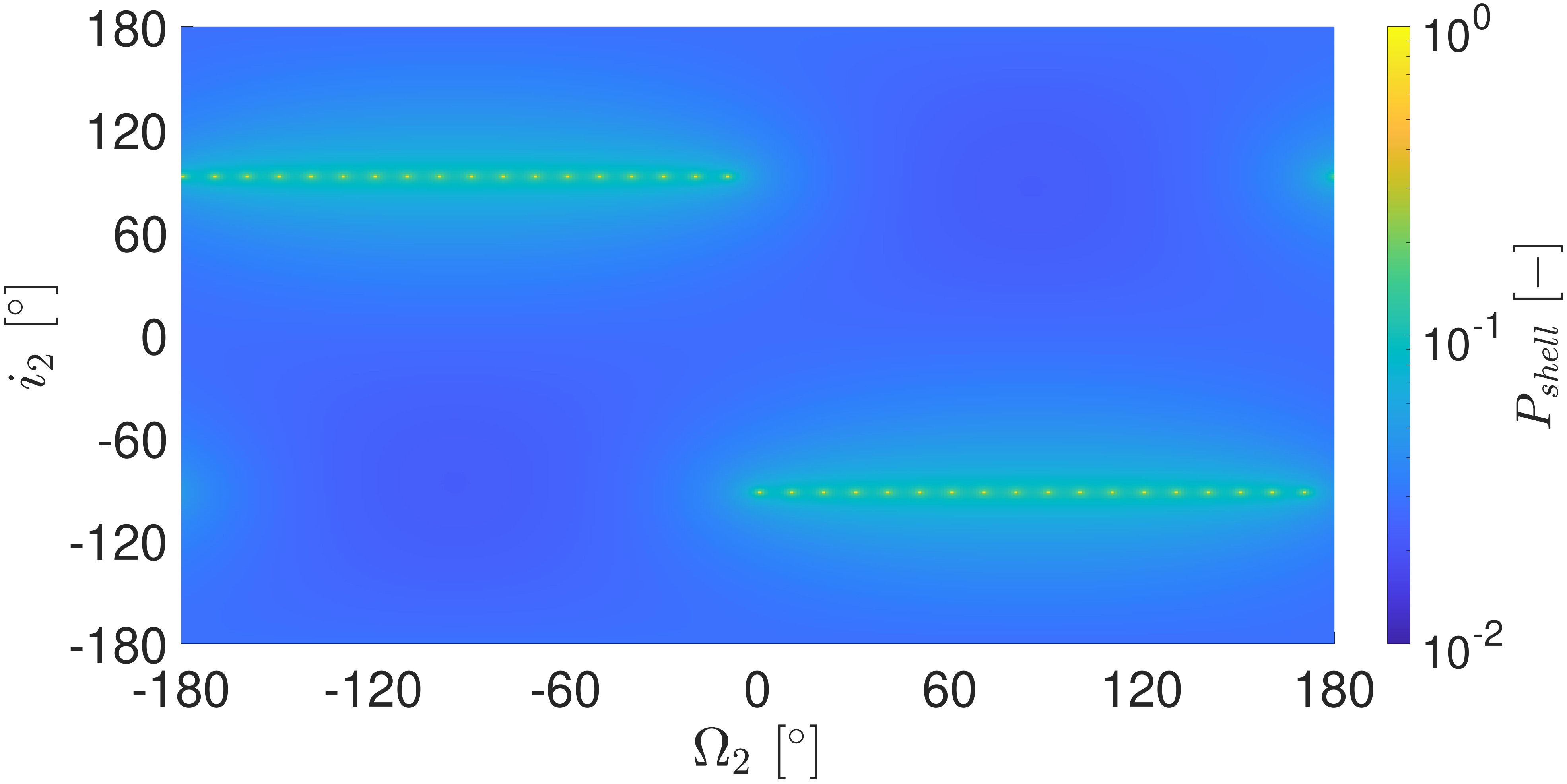}
    \caption{Collision probability distribution of passive de-orbiting through OneWeb shell.}
    \label{ob_cross_distrib}
\end{figure}

\section{Constellation's replacement risk assessment}
\label{result1_sec}

The statistical model developed in Section \ref{model_sec} will be used in this section to appraise the environmental hazard of satellite constellations. In particular, a risk assessment analysis of the disposal and injection phases of satellite constellations will be performed, also considering the presence of space debris and different Post-Mission Disposal (PMD) success rates. \\
The constellations considered are summarised in Table \ref{catalogue} and they are all modeled as Walker constellations, which are fully described by the inclination $i_1$, the total number of satellites $N_T$, the number of orbital planes $N_P$, the relative phase shift between satellites on adjacent planes $f$ and the altitude of the shell $h$. Most of these constellations were approved by the FCC, the authority regulating telecommunications in the United States. \\
The current section investigates, from a collision probability standpoint, the environmental consequences of the increase in LEO traffic that would result from constellation's satellites that have to be replaced, once they have reached their EoL. \\

\subsection{Optimal crossing}

\begin{figure}[b!]
    \centering
    \includegraphics[width=.9\linewidth]{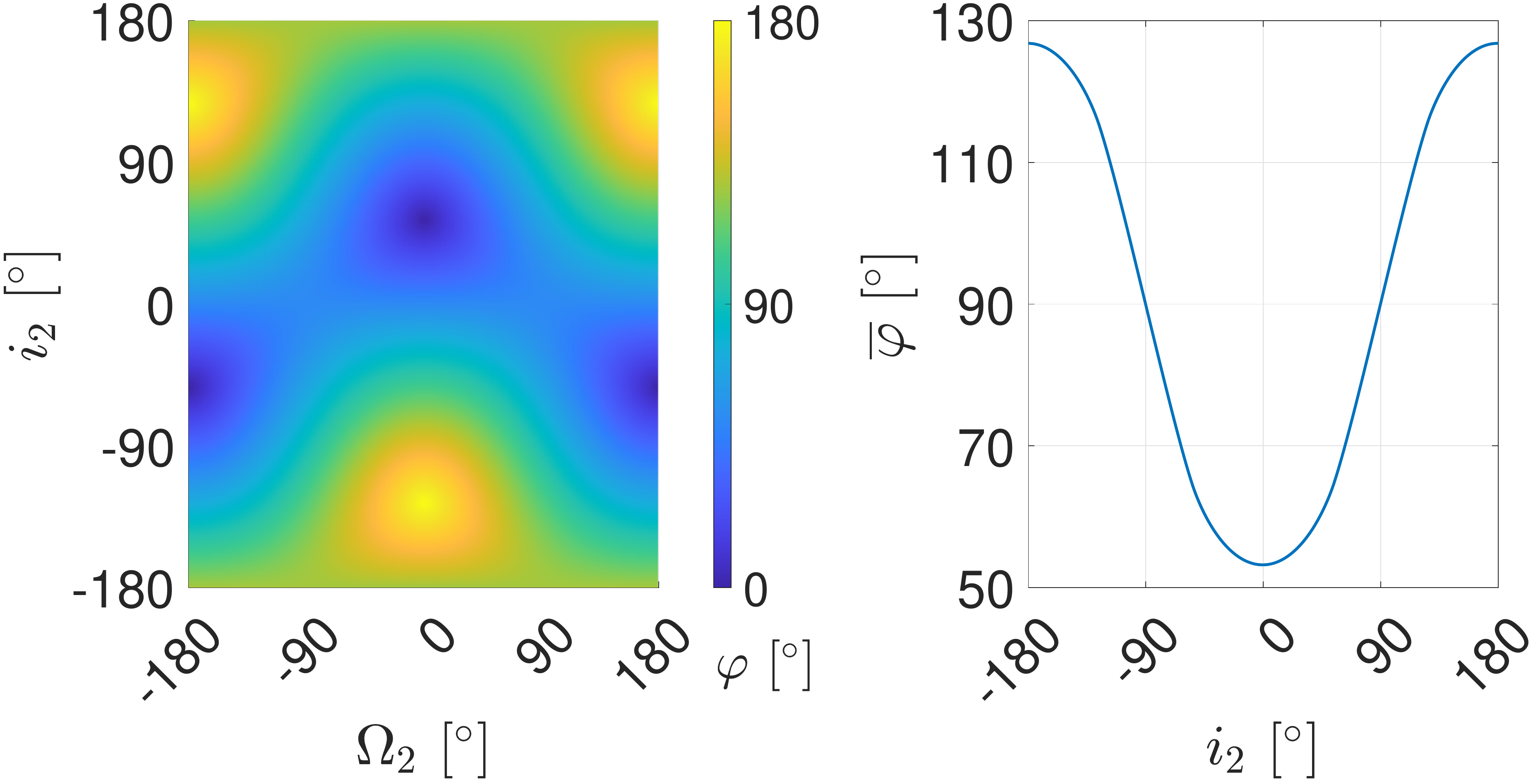}
    \caption{Distribution of the collision angle $\varphi$ between one orbital plane of Starlink (4) and any possible crossing orbital plane (left) and its mean value over the RAAN $\overline{\varphi}$ (right).}
    \label{s4_phi_distrib}
\end{figure}
\begin{figure}[b!]
    \centering
    \includegraphics[width=.9\linewidth]{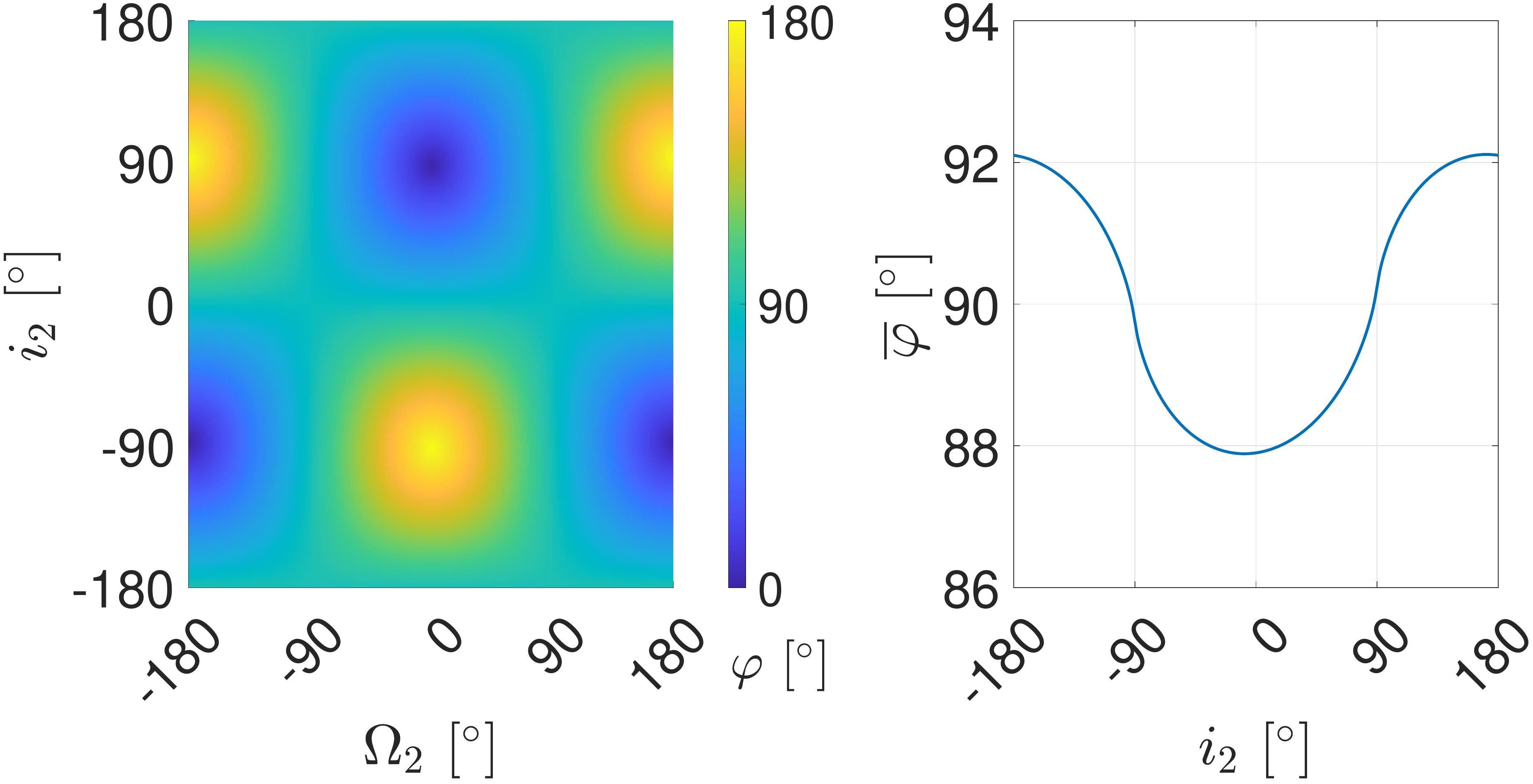}
    \caption{Distribution of the collision angle $\varphi$ between one orbital plane of OneWeb and any possible crossing orbital plane (left) and its mean value over the RAAN $\overline{\varphi}$ (right).}
    \label{ob_phi_distrib}
\end{figure}
The structure of a satellite constellation is considered to be fixed in time, thus the inclination $i_1$ and RAAN $\Omega_1$ of each orbital plane are constant. The relative orientation between constellation's and crossing satellites, hence, depends on the inclination and RAAN of the crossing satellite. \\
In Fig.s \ref{s4_cross_distrib} and \ref{ob_cross_distrib} are represented the collision probability distributions related to, respectively, a Starlink (4) and OneWeb  shell-crossing event, for different inclinations $i_2$ and RAANs $\Omega_2$ of the crossing satellite.
Each peak in Fig.s \ref{s4_cross_distrib} and \ref{ob_cross_distrib} corresponds to the case of head-on collision with one of the orbital planes of the constellation. It can also be noticed that the minimum mean collision probability $P_{shell}$ occurs when $i_2 = 0^\circ$. \\
To better understand this last statement, consider the collision angle distributions between a crossing satellite and a Starlink (4) and OneWeb orbital plane, respectively shown in Fig.s \ref{s4_phi_distrib} and \ref{ob_phi_distrib}. \\
The collision angle $\varphi$ between $\kappa_1$ and $\kappa_2$ can be computed as:
\begin{equation}
    \cos \varphi = \sin i_1 \sin i_2 \cos \Delta \Omega + \cos i_1 \cos i_2
    \label{cos_phi}
\end{equation}
Because $i_1$ and $i_2$ are constant, the distribution of collision angles $\varphi_i$ between the crossing orbital plane and the various constellation's orbital planes depends on $\cos(\Delta \Omega_i)$. As the constellation's orbital planes are equally spaced in RAAN, the collision angles characterising a shell-crossing event follow a cosinusoidal-like behaviour, whose mean value $\overline{\varphi}$ can be computed as:
\begin{equation}
    \cos(\overline{\varphi}) = \cos i_1 \cos i_2
    \label{varphi_mean}
\end{equation}
Imposing the derivative of Eq. (\ref{varphi_mean}) over $i_2$ equal to zero, returns the condition for which $\overline{\varphi}$ is either maximum or minimum. That is when $i_2$ is either null or straight. In particular, $\overline{\varphi}$ is minimum if the crossing orbital plane has inclination $i_2$ such that:
\begin{subequations}
\begin{align}
    i_2 & = 0^{\circ}  & \textrm{if} \ \cos i_1 \geq 0 \ \ \ \ \ \ \ \ \ \ \ \ \ \ \ \ \ \ \ \ \ \ \ \ \ \ \ \ \ \ \ \ \ \ \
    \label{oc1}
    \\
     i_2 & = 180^{\circ}  & \textrm{if} \ \cos i_1 \leq 0 \ \ \ \ \ \ \ \ \ \ \ \ \ \ \ \ \ \ \ \ \ \ \ \ \ \ \ \ \ \ \ \ \ \ \
     \label{oc2}
\end{align}
\label{optima_cross}%
\end{subequations}
Substituting Eq.s (\ref{oc1}) and (\ref{oc2}) into Eq. (\ref{cos_phi}), one obtains that $\cos \varphi$ is constant for all $i$, and always greater than zero. This is a rather important condition for collision probability minimisation, as the collision angles are guaranteed to be lower than (or equal to) $90^\circ$.
\begin{table*}[t!]
\centering
\caption{Data used in Section \ref{result1_sec}.}
\begin{tabular}{|l|l|l|l|}
\hline

& Value & Symbol \\ \hline

$S_1$ covariance & $\textrm{diag} \{ 0.25,1,0.25 \} \ \textrm{km}^2$ & $\boldsymbol{\zeta}_1^{RSW}$ \\ \hline

$S_2$ covariance & $\textrm{diag} \{ 1,4,1 \} \ \textrm{km}^2$ & $\boldsymbol{\zeta}_2^{RSW}$ \\ \hline

Drag coefficient & $2.2$ & $C_D$ \\ \hline

Input power & $600 \ \textrm{W}$ & $P_t$ \\ \hline

Thruster efficiency & $0.5$ & $\eta_t$ \\ \hline

Specific impulse & $2000 \ \textrm{s}$ & $I_{sp}$ \\ \hline

Maximum $\tfrac{a_1}{\sigma_\theta}$ ratio & $12.5$ & $\Phi_{max}$ \\

\hline
\end{tabular}
\label{data1_tab}
\end{table*}\\
Up to this point, one could argue that the results obtained in Eq. (\ref{optima_cross}) minimise the average collision angle $\overline{\varphi}$, but not necessarily the collision probability of the shell-crossing event $P_{shell}$. Nonetheless, $P_{shell}$ is characterised by a super-linear dependence on $\varphi$, meaning that, the larger the amplitude of the distribution of collision angles $\varphi_i$, the larger the overall collision probability $P_{shell}$. The conditions expressed in Eq. (\ref{optima_cross}) guarantee a constant distribution of collision angles $\varphi_i$ with minimum mean value, thus minimising $P_{shell}$. \\
Considering that the inclinations in Table \ref{catalogue} are, for the most part, lower than $90^\circ$. then, the equatorial plane would be the optimal gateway orbital plane for both satellite injection and disposal for collision probability minimisation. A side advantage of the optimal crossing orbit is that it has the optimal inclination needed in order to fully exploit Earth rotation during launch, hence reducing costs. \\
It must be highlighted that the result obtained for optimal crossing is limited to constellations whose orbital planes have RAAN spanning over $360^\circ$. Indeed, this is not the case for constellations characterised by polar orbits, for which RAAN interval is typically less than $180^\circ$. This feature allows for shell-crossing events with collision angles always smaller than (or equal to) $\overline{\varphi}$, as defined in Eq. (\ref{varphi_mean}).\\
Consider the $18$ orbital planes of OneWeb constellation to have equally spaced RAANs between $0^\circ$ and $170^\circ$ and inclination $i_1 = 87.9^\circ$. Then, optimal crossing can be achieved with a crossing orbital plane having inclination $i_2$ equal to $i_1$, and RAAN equal to $85^\circ$. Nonetheless, repeatedly changing the inclination of the crossing satellite from $0^\circ$ to near-polar and vice versa is broadly inefficient. Moreover, crossing polar constellation through the equatorial plane leads to constant collision angles marginally lower than (or equal to) $90^\circ$, and, as it can be seen from Table \ref{table_validation}, $P_{shell}$ is less affected at low $\varphi$. \\
Finally, optimal crossing can, sometimes, be a poor choice in terms of fuel (or time) consumption. Furthermore, as the changes in $P_{shell}$ are considerable only for large values of collision angles, trade-offs between collision probability minimisation and fuel consumption are, in general, recommended.

\subsection{Methodology}
\label{const_rep_meth}

The risk assessment is performed by considering the EoL satellite that has to de-orbit and the new satellite that shall replace it. For both satellites, space debris and constellations that have to be crossed are considered as collision sources and the overall collision probability is computed as a combination of these two. The collision probability with other constellations is assessed using the model developed in Section \ref{model_sec}, whereas the ESA-MASTER software, as done in \citep{bib_radtke} and \citet{bib_lemay}, is used for the evaluation of the collision probability with space debris \citep{bib_tesi}. \\
Let $P^{\pm}$ be the collision probabilities of the injecting and de-orbiting satellite, respectively; then:
\begin{equation}
    P^{\pm} = 1 - (1 - P_c^{\pm})(1 - P_d^{\pm})
\end{equation}
where $P_c$ and $P_d$ are the probabilities that a collision will occur, respectively, with constellations and debris. \\
The overall collision probability related to the full replacement of a shell, can finally be computed as:
\begin{equation}
    P_{tot} = 1 - [(1 - P^+)(1 - P^-)]^{N_T}
\end{equation}
The constellations are eventually divided into two categories, based on their scope: Telecommunication (T) or Earth-Observation (EO). Satellites belonging to either category are modeled as spheres, the former having an average radius of $2$ m and mass of $200$ kg, the latter of $0.5$ m and $50$ kg. The combined hard sphere radius $r_a$ is given by the sum of the cross-sectional radii of constellation and crossing satellites. Table \ref{data1_tab} collects data used for simulations.

\subsection{Results}

\begin{figure}[b!]
    \centering
    \includegraphics[width=\linewidth]{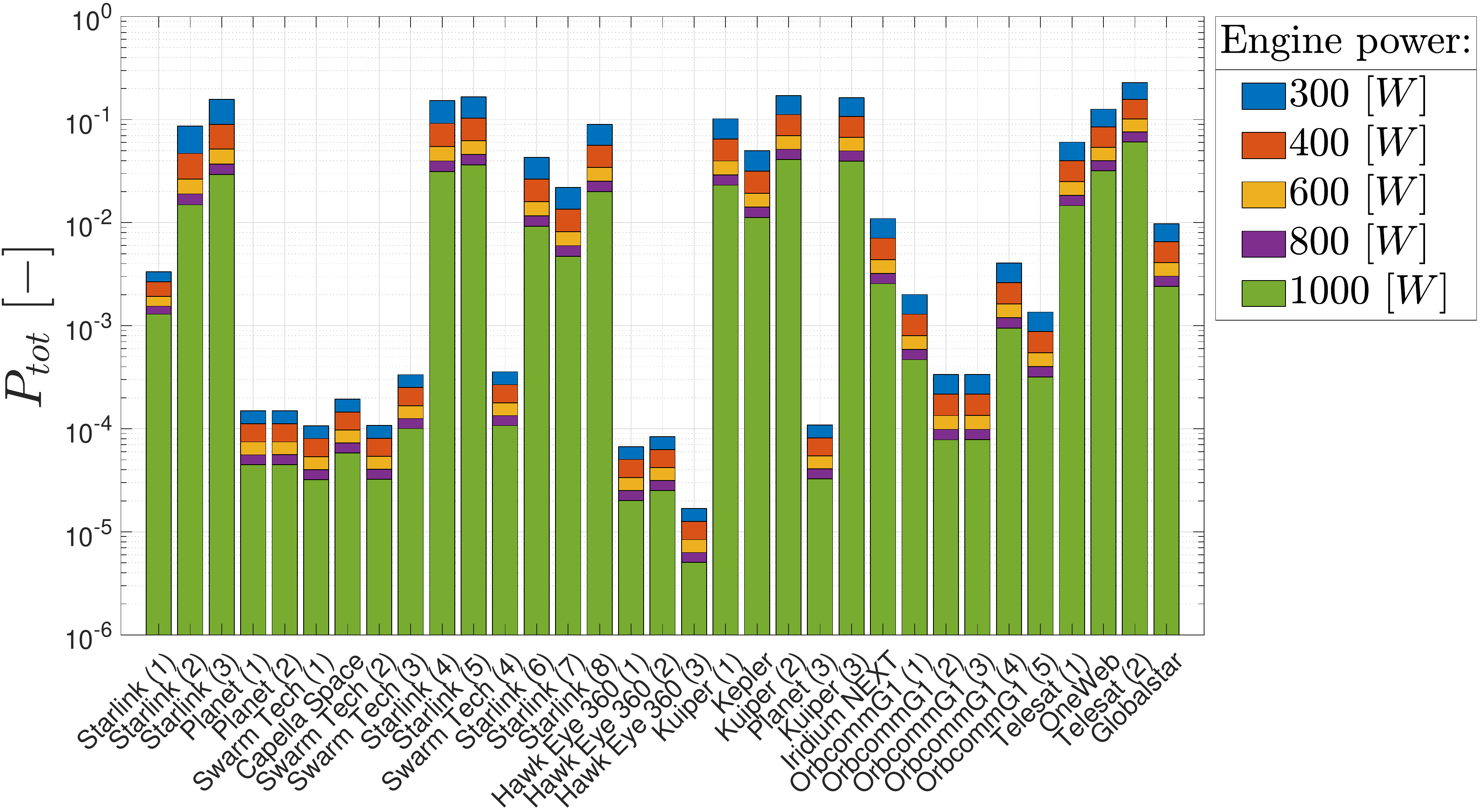}
    \caption{Collision probability related to the replacement of the whole constellation at different engine input powers considering scenario \textbf{B}.}
    \label{repl1}
\end{figure}

\begin{figure}[b!]
    \centering
    \includegraphics[width=\linewidth]{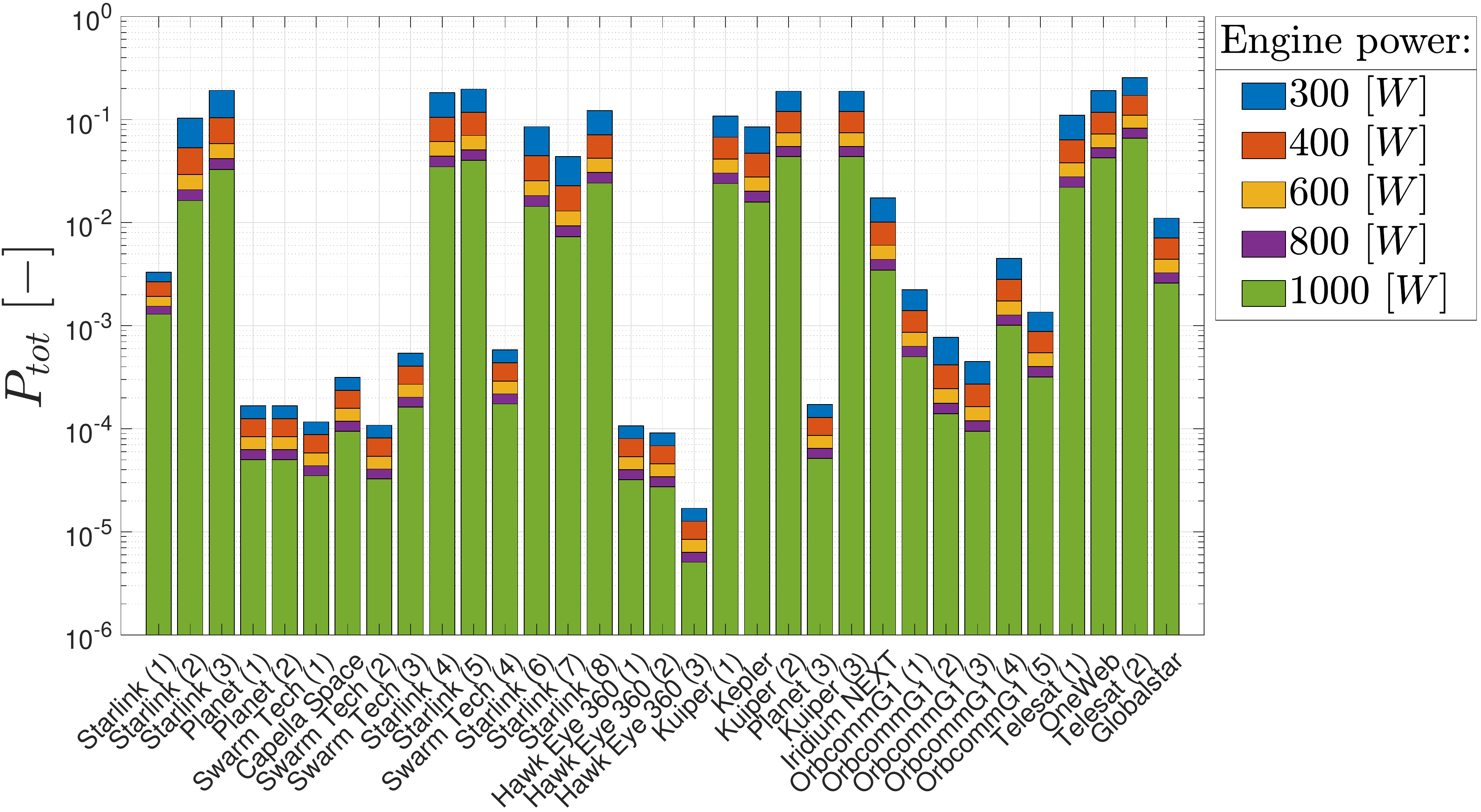}
    \caption{Collision probability related to the replacement of the whole constellation at different engine input powers considering scenario \textbf{N}.}
    \label{repl2}
\end{figure}

The collision probability of a full constellation replacement has been evaluated for every constellation considering different engine input powers and two different scenarios: best case scenario (\textbf{B}) corresponds to a crossing orbit having zero inclination, whereas, in nominal case scenario (\textbf{N}) the crossing orbital plane is assumed to be an orbital plane of the arrival or departure shell. It follows that the inclination of the crossing satellite will be $i_2 = i_1$. Results are shown in Fig.s \ref{repl1} and \ref{repl2}.
\begin{figure}
\centering
    \centering
    \includegraphics[width=.94\linewidth]{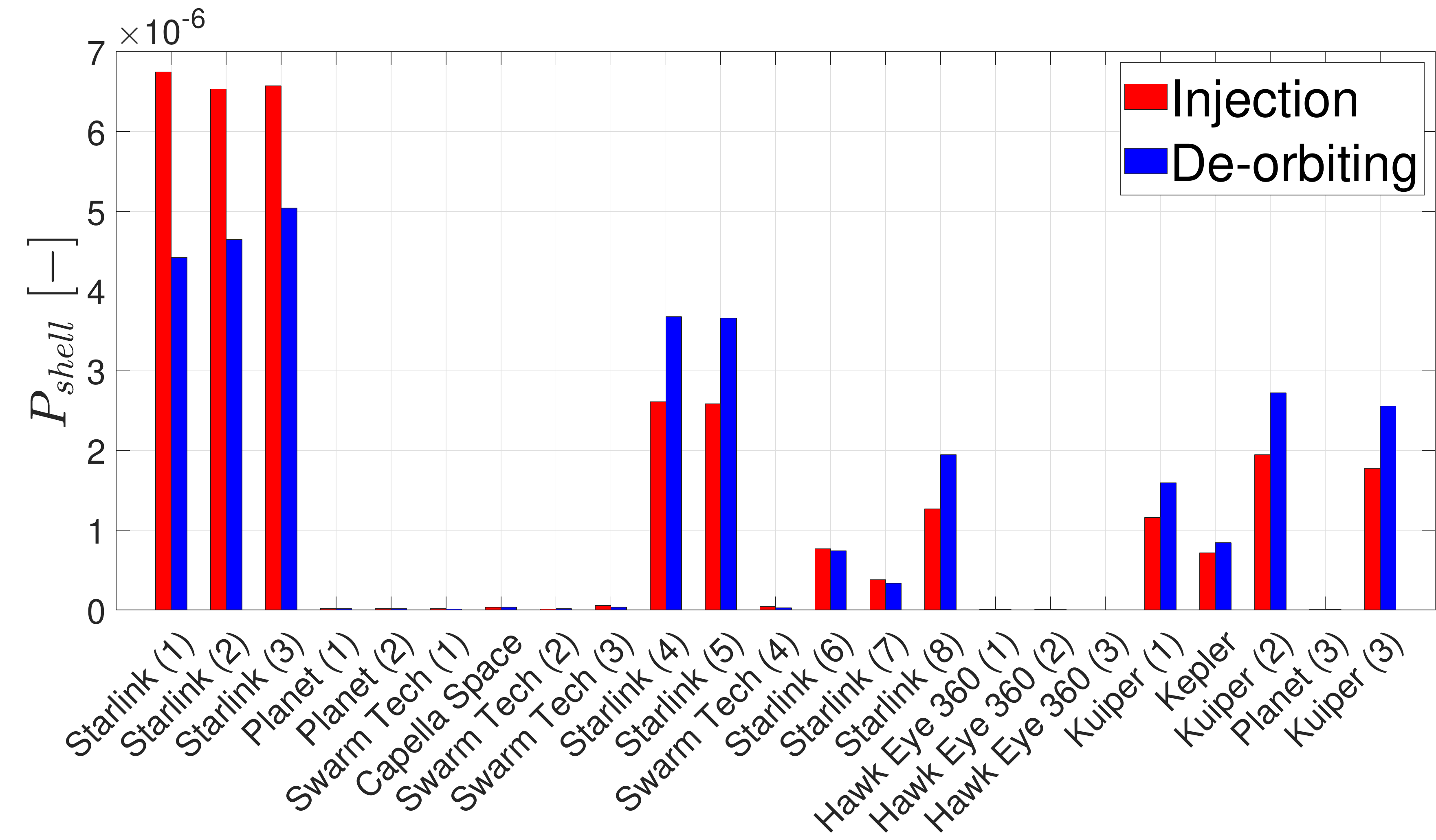}
    \caption{Collision probability with each lower altitude constellation of one injecting and de-orbiting Kuiper (3) satellite with an engine input power of $600$ W.}
    \label{Rs1}
\end{figure}
\begin{figure}
\centering
    \includegraphics[width=.96\linewidth]{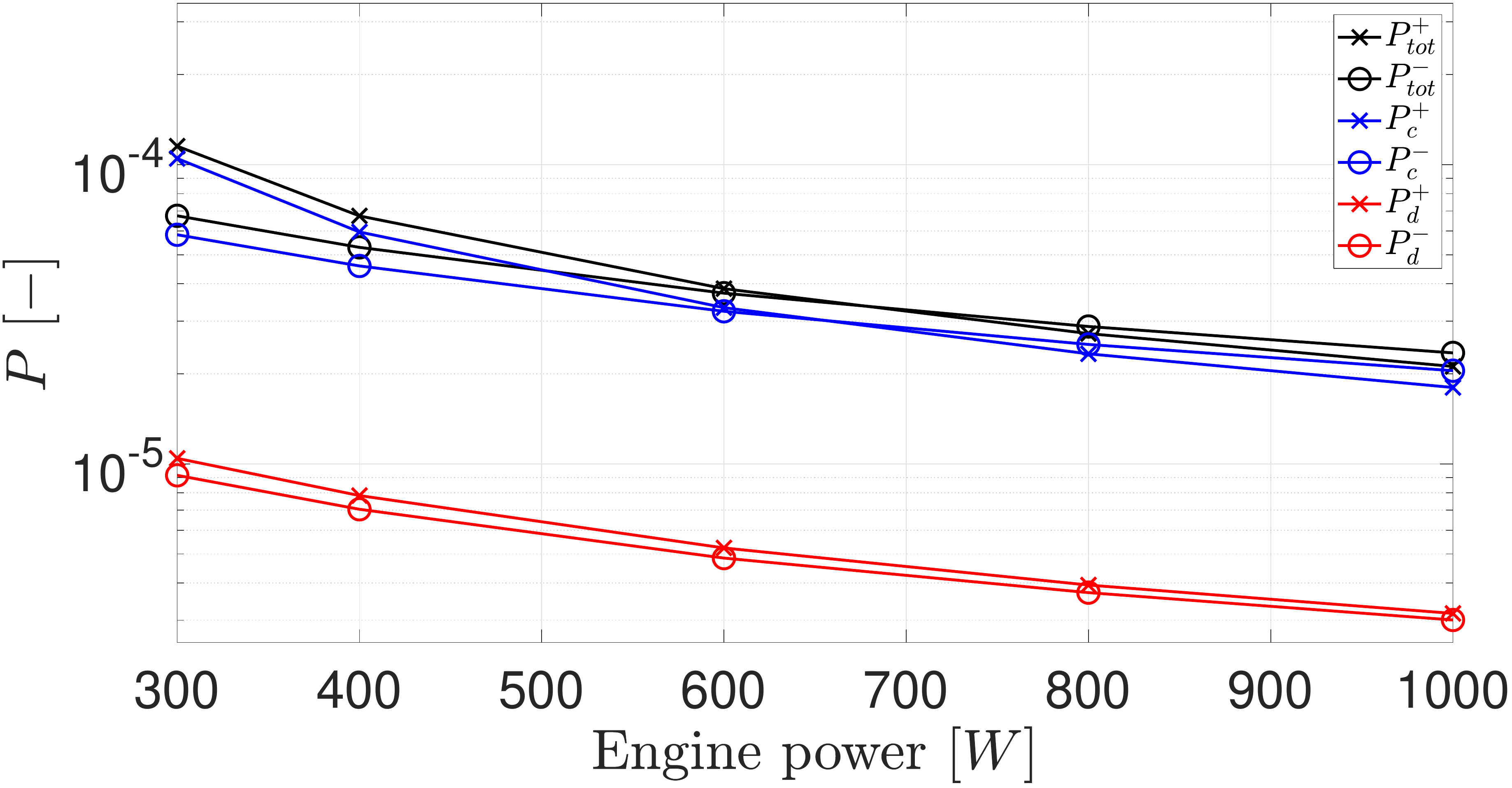}
    \caption{Collision probability per source and engine input power of one injecting and de-orbiting Kuiper (3) satellite. Optimal injection and nominal de-orbiting have been considered.}
    \label{Rs2}
\end{figure}
\begin{figure}
    \centering
    \includegraphics[width=\linewidth]{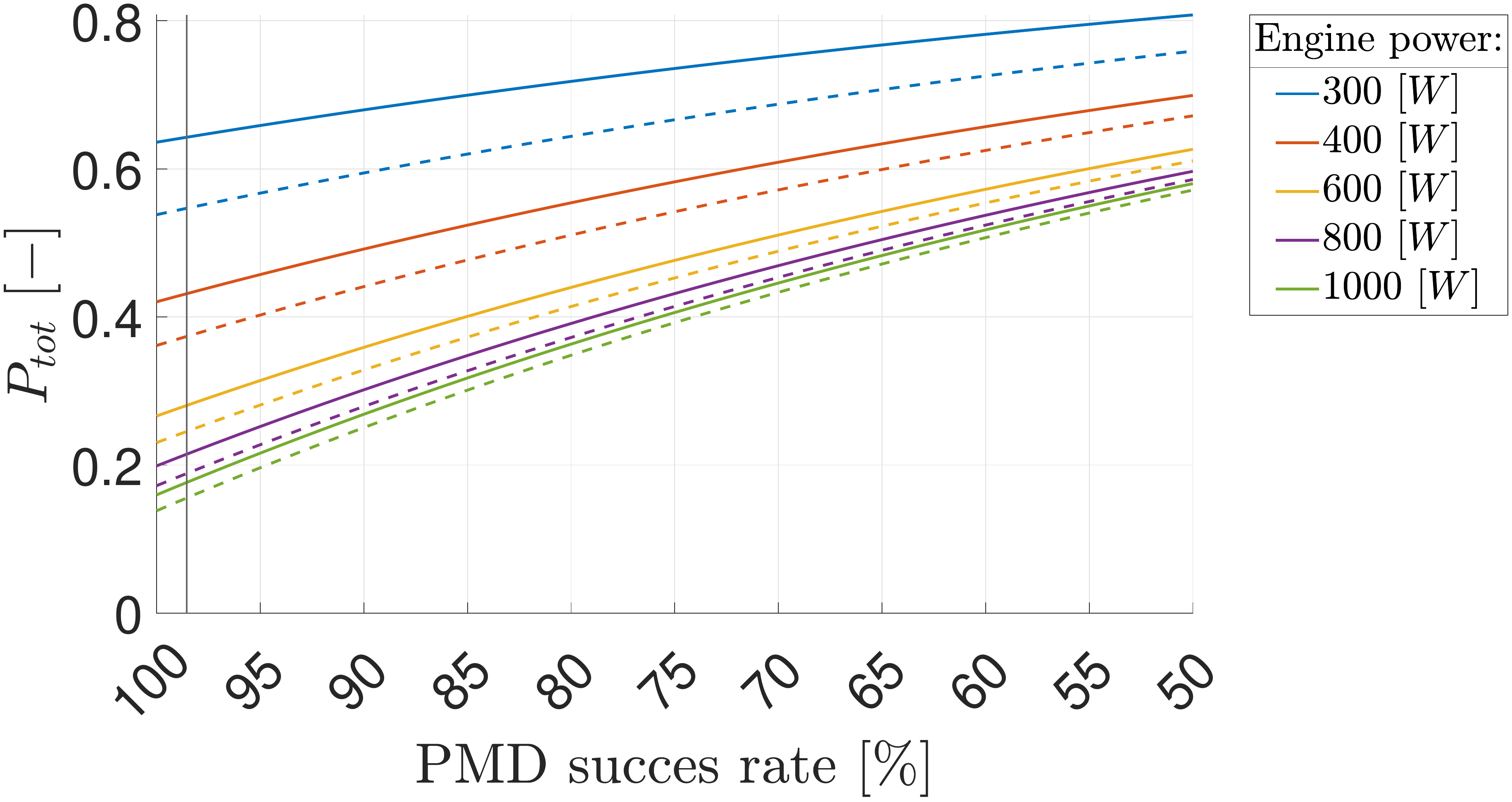}
    \caption{Overall collision probability related to the replacement of all $11926$ Starlink satellites at different PMD success rates considering both scenarios \textbf{N} (continuous) and \textbf{B} (dashed).}
    \label{starlink_pmd}
\end{figure}\\
Fig.s \ref{Rs1} and \ref{Rs2} show more details about the risk assessment related to the replacement of Kuiper ($3$) shell, considering optimal crossing only for injecting satellites. Recalling Section \ref{const_rep_meth}, in Fig. \ref{Rs1} are represented the collision probabilities $P_c^{\pm}$, $P_d^{\pm}$ and $P^{\pm}$ for different values of engine input power, and Fig. \ref{Rs2} shows how $P_c^{\pm}$ are partitioned among the various shells. \\
Finally, by assuming that failed EoL satellites re-enter through passive de-orbiting, different PMD success rates can be considered for the risk assessment analysis. As the Starlink constellation is, by far, the most populated one, in Fig. \ref{starlink_pmd} are shown the collision probabilities related to the full replacement of the Starlink constellation with different PMD success rates, considering both scenarios \textbf{B} and \textbf{N}. The vertical black line refers to a failure rate of $1.45\times10^{-2}$, which is the one currently declared by Starlink \citep{bib_kramer}.\\
As expected from the results collected in Table \ref{table_validation}, Fig.s \ref{repl1} and \ref{repl2} show that the improvement of scenario $\boldsymbol{\textrm{B}}$ over $\boldsymbol{\textrm{N}}$, are not considerable. The benefits of optimal crossing can be appreciated in Fig. \ref{starlink_pmd}, where a large number of satellites is considered. \\
The effects of the thruster acceleration on the collision probability can be observed in Fig.s \ref{repl1} and \ref{repl2}, where it is shown that increasing the propulsion of the satellite can reduce it by one order of magnitude at higher altitude. \\
Other parameters that affect replacement collision probability are size and number of constellation satellites. Due to their much smaller size, Earth observation satellites are characterised by impact probabilities several orders of magnitude lower than telecommunication satellites. A similar, but less marked behaviour, can be noted by comparing crowded and uncrowded telecommunication shells. \\
Besides the benefits of a powerful thruster, it can be noticed from Fig. \ref{starlink_pmd} that the PMD declared by Starlink does not substantially change the risk assessment with respect to the case of $100$ \% PMD succes rate. Nevertheless, it must be emphasised that collisions with satellites belonging to the departure or arrival shell are not contemplated. \\
Considering Fig. \ref{Rs1}, a satellite on its journey has a higher chance of colliding with a constellation satellite, rather than with a debris. On the other hand, constellation satellites can be tracked much more easily and, unlike debris, are manoeuvrable. \\
By looking at Fig. \ref{Rs2} it is also clear that the Starlink shells in Very-Low Earth Orbit (VLEO) pose the greatest threat. The large gap between injection and de-orbiting collision probabilities is due to the low altitude of the shells, at which the atmospheric drag is considerable, as drag accelerates de-orbiting while slowing down injection. Moreover, by looking at Fig.s \ref{repl1} and \ref{repl2}, the collision probability for the replacement of Starlink VLEO shells is comparable to higher altitude shells, even if they have to go through far fewer shells. The reason for this is the large number of satellites that VLEO shells require to ensure (near) global coverage, thus creating a thin region of space heavily populated. On the other hand, it must be emphasised that the lower the altitude of the shell, the faster the re-entry of the debris generated from an in-orbit collision. \\
A final remark shall be made about the size of the satellites. Indeed, the collision probabilities considering spherical shapes are, in general, larger than the actual values. This issue is treated in \citet{bib_lemay}, where it is shown that the average number of collisions between space debris and one Starlink satellite modeled as a sphere, is around $8$ \% larger than the one computed using a box wind model.

\section{Preliminary analysis of the consequences of an in-orbit catastrophic collision}
\label{result2_sec}

In the following, a risk assessment related to the interaction between satellite constellations and a cloud of fragments, generated by an in-orbit catastrophic collision, will be performed. The model developed in Section \ref{model_sec} will be used for the evaluation of the collision probability between each generated fragment that, due to its passive de-orbiting, has to cross lower altitude satellite constellations.

\subsection{Methodology}

NASA standard breakup model, described in \citet{bib_johnson}, has been utilised for the evaluation of the area-to-mass ratio A/M and area $A$ of the fragments, having characteristic lengths between $2$ cm and $2$ m. This interval of characteristic lengths has been divided into $100$ logarithmic bins for the evaluation of the total number of fragments $N_f$, as done similarly in \citet{bib_letizia}. Assuming spherical shaped fragments, both the radius $R$ and the mass $M$ can also be determined. \\
As discussed in \citet{bib_jehn}, \citet{bib_ashenberg} and \citet{bib_letiziacolombo}, Earth's oblateness spreads the fragments along the argument of periapsis $\omega$ and RAAN $\Omega$. At this point, again according to \cite{bib_letiziacolombo}, the Keplerian parameters can be randomised and the only perturbing acceleration that has to be accounted for the propagation of the fragments is atmospheric drag. \\
In this study, the risk assessment of every shell-crossing event is performed by assuming that each fragment has an inclination $i_2$ equal to the one of the satellite they have been generated from, and RAAN $\Omega_2$ which is uniformly randomised between $0^{\circ}$ and $360^{\circ}$. \\
Furthermore, the collision angles $\varphi_i$ between a crossing object and each constellation's orbital plane are quite spread, since they are equally spaced along the RAAN. The randomisation of the collision angles would, therefore, provide good statistical results if used for hundreds of thousands of shell-crossing events, as it is the case for the risk assessment of a debris cloud crossing a shell. \\
The covariance matrices of the fragments are modeled as diagonal in their respective RSW frames, with fixed axes ratios and randomised determinant:
\begin{equation}
    \boldsymbol{\zeta}^{RSW} = \rho \begin{bmatrix}
    1 & 0 & 0 \\
    0 & 4 & 0 \\
    0 & 0 & 1 \\
    \end{bmatrix} \ \ \ , \ \ \  5 \leq \rho \leq 10
\end{equation}
It was observed that neglecting the covariance matrices and using Eq. (\ref{final3_eq}) produces the same results, since for most orbit orientations, the condition $\varphi \leq \varphi^*$ holds. \\
Depending on the physical properties of fragments and constellations' satellites, it can be determined whether an eventual collision would be catastrophic or non-catatrophic, and the debris trackable or non-trackable. The former distinction is solved using a statistical approach \citep{bib_tesi}. Fragments with mass equal or greater than a threshold value $M_p^*$ are considered to cause a catastrophic collision. The critical mass is defined as:
\begin{equation}
    M_p^* = \frac{E_k M_1 a_1}{\mu(1-\cos i_1 \cos i_2)}
    \label{cat_mass}
\end{equation}
where $M_1$ is the mass of the target satellite and $E_k = 40$ $\frac{\text{J}}{\text{g}}$ is the minimum impact energy per target mass necessary for a collision to be catastrophic \citep{bib_krisko}. A length of $10$ cm \citep{bib_radtke} has been chosen as a threshold value between trackable and non-trackable objects. \\
Finally, in \citet{bib_letizia} a simulation was carried of a non-catastrophic collision characterised by a impact energy of $50$ kJ. It was shown that almost every fragment has an eccentricity between $0$ and $3\times10^{-1}$, half of which are lower than $5\times10^{-2}$. For this reason, the results collected in Section \ref{coll_res} are yet to be completely validated.

\subsection{Results}
\label{coll_res}

\begin{figure}[t!]
    \centering
    \includegraphics[width=\linewidth]{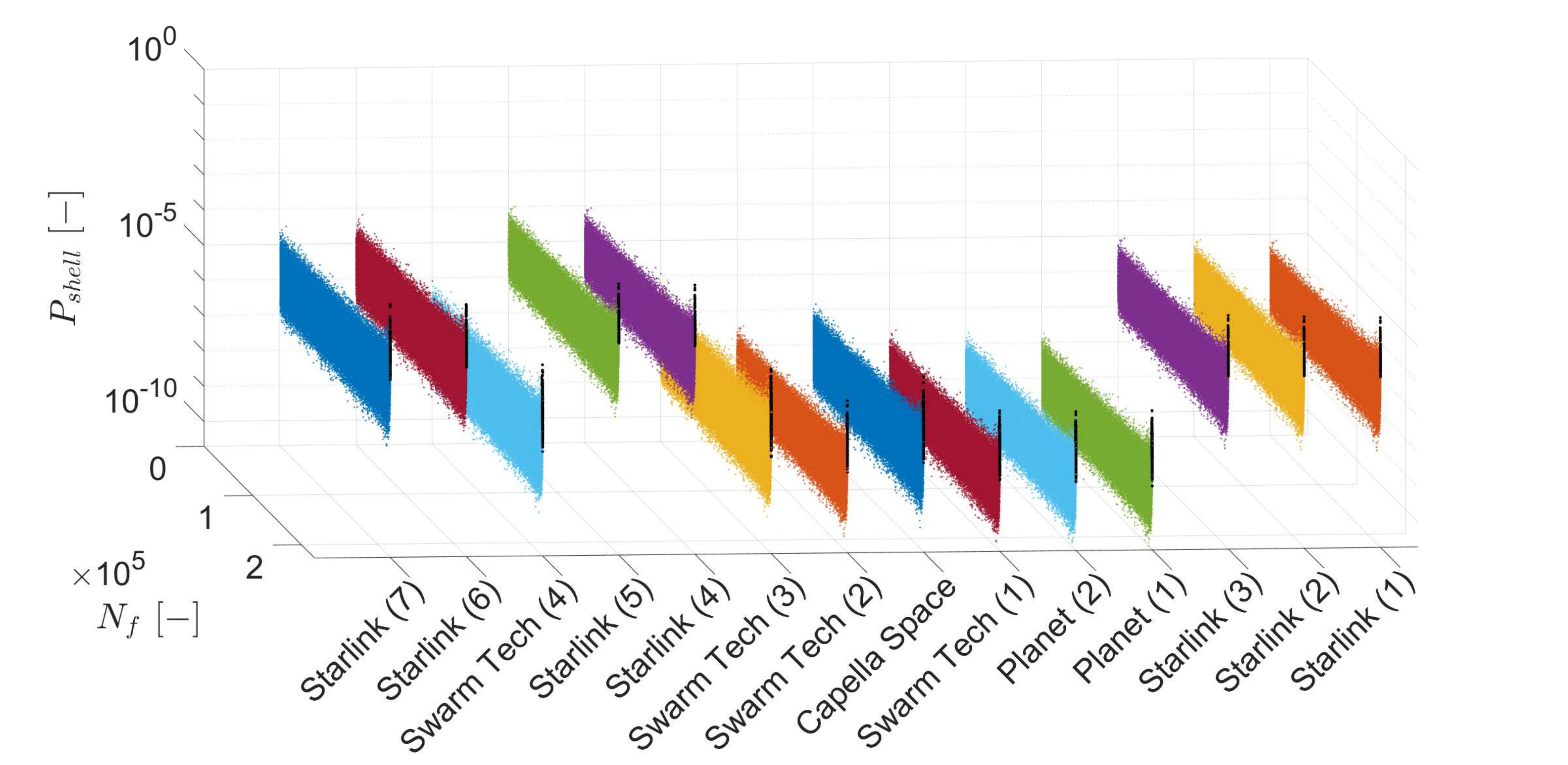}
    \caption{Collision probabilities of the Starlink-debris collision cloud per shell and fragment. Fragments are sorted based on their characteristic lengths.}
    \label{sd2}
\end{figure}

\begin{figure}[t!]
    \centering
    \includegraphics[width=\linewidth]{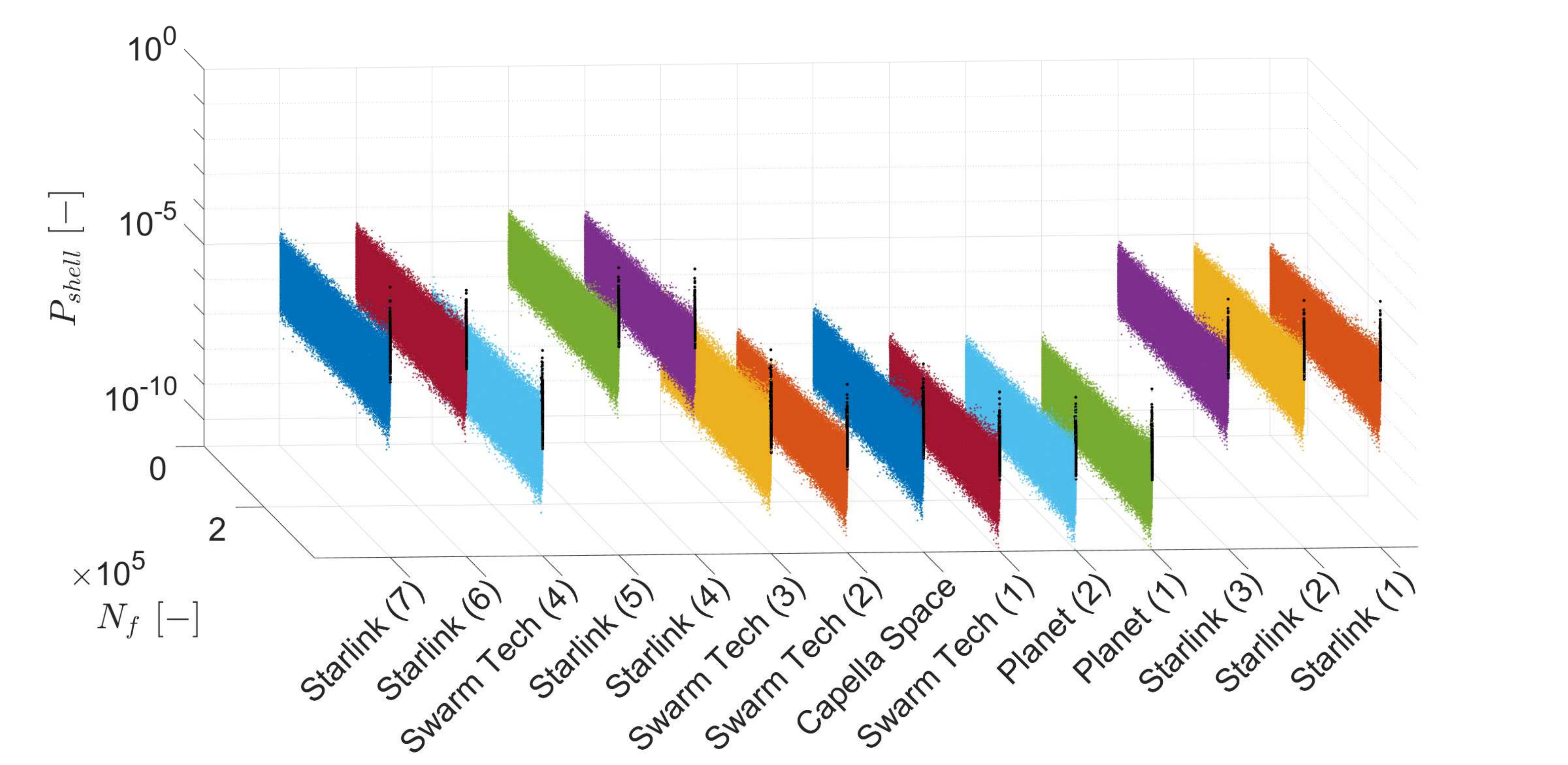}
    \caption{Collision probabilities of the Starlink-Starlink collision cloud per shell and fragment. Fragments are sorted based on their characteristic lengths.}
    \label{ss2}
\end{figure}

\begin{figure}[t]
    \centering
    \includegraphics[width=\linewidth]{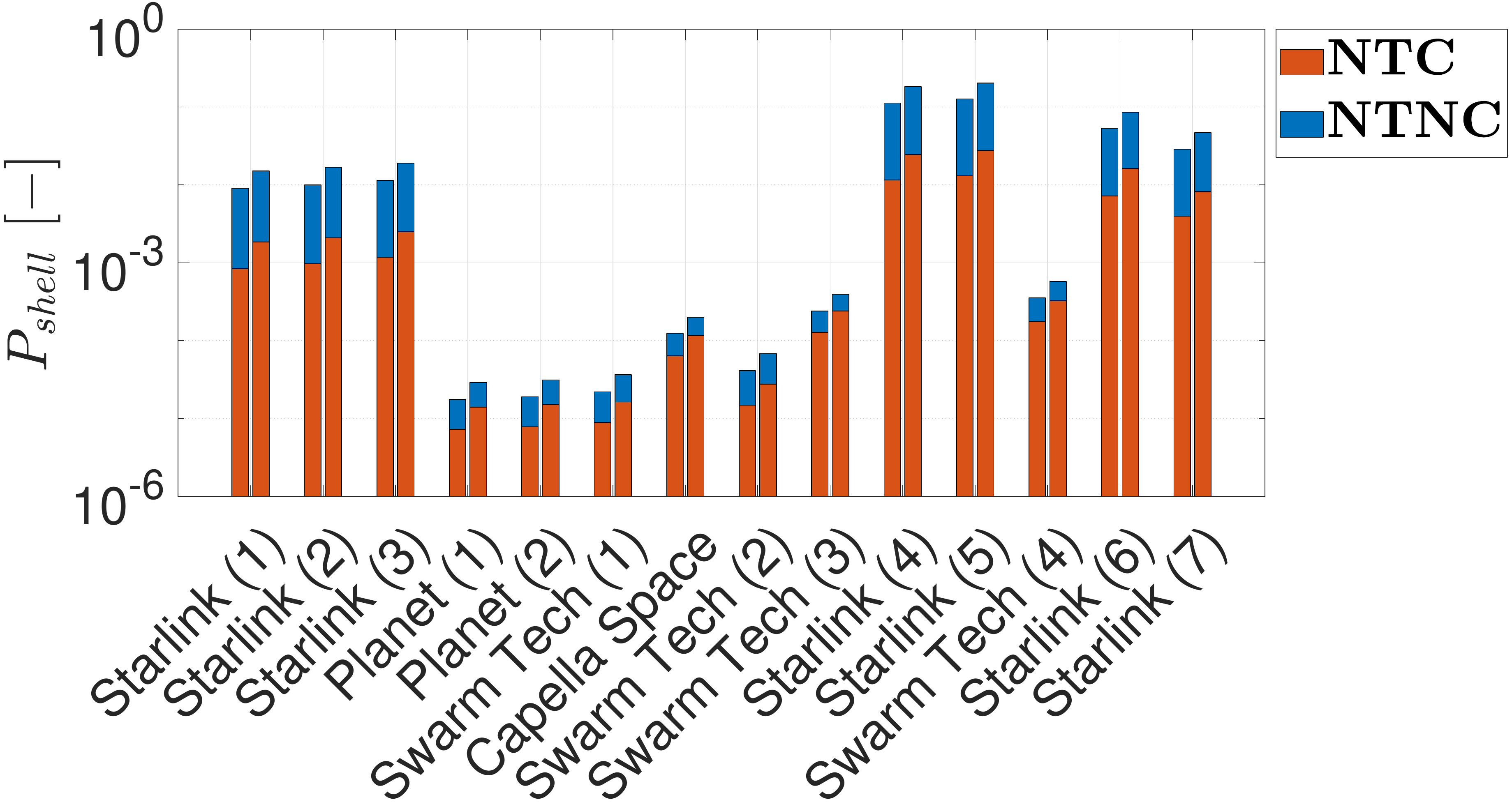}
    \caption{Collision probability per shell of non-trackable catastrophic (\textbf{NTC}) and non-trackable non-catastrophic (\textbf{NTNC}) collisions considering fragments from Starlink-debris (left) and Starlink-Starlink (right) collision.}
    \label{ntncntc}
\end{figure}

Two different scenarios have been considered: a catastrophic collision between a $200$ kg Starlink (8) satellite and a $10$ kg debris; a catastrophic collision between two $200$ kg Starlink (8) satellites. \\
The collision probabilities between each fragment and each shell crossed are shown in Fig.s \ref{sd2} and \ref{ss2}. The black dots refer to catastrophic collision probabilities. These probabilities are in general larger than the non-catastrophic collision probabilities, as catastrophic collision are caused by heavy fragments, which, usually, have large cross-sectional areas.
\\
As collisions with trackable objects can be prevented much more easily, only results about collision probabilities with non-trackable debris are shown in Fig. \ref{ntncntc}.
\\
Non-trackable non-catastrophic collisions are the main threat posed by a fragmentation event. Although catastrophic collisions result in the total disruption of the target, small objects are, in general, more numerous and they would not only contribute to the generation of more fragments, but also damage the impacting satellite, thus increasing the failure rate. Due to their smaller mass, Earth observation satellites have a relatively higher chance of a non-trackable catastrophic collision. \\
The results obtained show that an in-orbit collision would most probably decree LEO inoperability due to cascade effect. Nonetheless, the dynamics involved in this type of shell-crossing events are, in general, very slow, thus granting a large time window to operate. Indeed, the higher the altitude at which the catastrophic collision may occurs, the slower the dynamics of the satellite, the easier preventing collisions. On the other hand, the higher the collision altitude, the longer the time needed for the disposal of the debris cloud. \\
Finally, to highlight the performances of the model, using a single core of an i7-4720 at $2.6$ GHz with $8$ GB of RAM, the risk assessments of the Starlink-debris and Starlink-Starlink collision scenarios required MATLAB $20.57$ and $33.55$ seconds, respectively. In this time windows, a total of $3,184,398$ and $5,163,074$ shell-crossing events were analysed.  \\

\section{Conclusion}
\label{conclusion_sec}

This study has proposed an analytical model for the assessment of the environmental threat posed by large satellite constellations, through a statistical modeling of the collision probability related to a shell-crossing event. The main advantage of the statistical model is the low computational effort, as the collision probability related to millions of shell-crossing events can be assessed within seconds on an average personal computer. The model was used to analyse the constellations replacement traffic and the consequences of an in-orbit catastrophic collision. \\
Constellation's satellites have to be replaced once they reach their EoL, thus generating an intense traffic in LEO. Both injection and disposal have been investigated, showing that, besides phasing, the main parameters of interest are the semi-major axis rate of change $|\Delta a|$ and the relative orientations between the constellation's and crossing orbital planes, summarised by the collision angle $\varphi$. As $|\Delta a|$ can be tuned through the proper selection of the satellite's thruster, a great emphasis was placed in the design of an optimal crossing orbit, which was found to be at zero inclination. The establishment of an international gateway orbital corridor for both injection and disposal might simplify the management of future space traffic scenarios, as satellites on the same orbits share the lowest possible probability that a collision will occur. \\
In the context of constellation replacements, an analysis of the risk assessment based on different PMD success rates has also been performed regarding the Starlink constellation. The results proved that the benefits arising from the selection of a powerful thruster are much greater than the ones arising from the decrease of the failure rate, proving once again that the main parameters of interest for the design of a shell-crossing manoeuvre is the semi-major axis rate of change. \\
The statistical model has also been used for the risk assessment of the debris flux generated from an in-orbit catastrophic collision. With emphasis on the probability of a collision with non-trackable objects, it was shown that, considered the presence of thousands of constellation's satellites, managing the consequences of a collision is definitely an arduous task. \\
Although the statistical model presented in Section \ref{model_sec} was developed based on the assumption of circular orbits, it was shown that the mean collision probability can be predicted with great accuracy for crossing orbit eccentricity $e_2$ up to $10^{-3}$. More elliptical crossing orbits are yet to be tested. It was also shown that low eccentricity crossing orbits allow for a greater collision probability minimisation, through proper phasing between satellites. \\
The main limitation of the model is that its reliability is not proven when the effects of $J_2$ perturbations are considered. Furthermore, the model is specifically designed for assessments of traffic scenarios and preliminary mission analysis, as the result is an estimation of the average collision probability, whereas the propagation method showed that the true anomaly phase between constellation's and crossing satellite is a parameter of great interest for the design of shell-crossing events. \\
The analysis of the limitations of this study has already suggested the main topics for possible future improvements. Indeed, the main focus would be to test the accuracy of the model in case of more elliptical crossing orbits, also including $J_2$ in the analysis. Another possibility would be to develop a semi-analytical model with the possibility of using covariance matrices not necessarily diagonal in their respective RSW frames. Finally, it is possible to analyse the collision probability profile $P(\delta \omega^{\{0\}})$ of a shell-crossing event in the frequency domain. This might be a convenient choice for the development of an analytical model, able to retrieve the collision probability profile based on the main parameters of the shell-crossing event, thus avoiding numerical methods exploitation.

\section{Aknowledgements}

This project has received funding from the European Research Council (ERC) under the European Union’s Horizon 2020 research and innovation programme [grant agreement No 679086 – COMPASS].

\end{document}